\documentclass[twocolumn]{aastex631}

\usepackage{comment}
\usepackage{amsmath}
\usepackage{mathrsfs}
\usepackage{graphicx}
\usepackage{subfigure}

\begin{document}




\title{Stellar Abundances as Probes of Rocky Exoplanet Interiors: The Mantle Composition and Mineralogy of GJ 486b}

\correspondingauthor{Liton Majumdar}
\email{liton@niser.ac.in, dr.liton.majumdar@gmail.com}

\author [0009-0000-2964-9450] {Harshit Krishna}
\altaffiliation{These authors contributed equally to this work.}
\affiliation{Exoplanets and Planetary Formation Group, School of Earth and Planetary Sciences, National Institute of Science Education and Research, Jatni 752050, Odisha, India}
\affiliation{Homi Bhabha National Institute, Training School Complex, Anushaktinagar, Mumbai 400094, India}
\affiliation{Department of Physical Sciences, Indian Institute of Science Education and Research (IISER) Mohali,
Knowledge City, Sector 81, Sahibzada Ajit Singh Nagar, Punjab 140306, India}

\author [0009-0005-0831-7744] {Asmit Gupta}
\altaffiliation{These authors contributed equally to this work.}
\affiliation{Exoplanets and Planetary Formation Group, School of Earth and Planetary Sciences, National Institute of Science Education and Research, Jatni 752050, Odisha, India}
\affiliation{Homi Bhabha National Institute, Training School Complex, Anushaktinagar, Mumbai 400094, India}
\affiliation{Indian Institute of Science Education and Research, Kolkata, Mohanpur 741246, West Bengal, India }

\author[0009-0003-2617-3938]{Sneha Bhowmik }
\affiliation{Exoplanets and Planetary Formation Group, School of Earth and Planetary Sciences, National Institute of Science Education and Research, Jatni 752050, Odisha, India}
\affiliation{Homi Bhabha National Institute, Training School Complex, Anushaktinagar, Mumbai 400094, India}
\affiliation{Department of Geology and Geophysics, Indian Institute of Technology, Kharagpur 721302, India}

\author[0000-0003-1057-2320]{Chandan K. Sahu}
\affiliation{Exoplanets and Planetary Formation Group, School of Earth and Planetary Sciences, National Institute of Science Education and Research, Jatni 752050, Odisha, India}
\affiliation{Homi Bhabha National Institute, Training School Complex, Anushaktinagar, Mumbai 400094, India}
\affiliation{Department of Geosciences, Pennsylvania State University, University Park, State College, 16802, Pennsylvania, USA}

\author [0009-0001-0265-7095] {Sudipta Mridha}
\affiliation{Exoplanets and Planetary Formation Group, School of Earth and Planetary Sciences, National Institute of Science Education and Research, Jatni 752050, Odisha, India}
\affiliation{Homi Bhabha National Institute, Training School Complex, Anushaktinagar, Mumbai 400094, India}

\author [0000-0001-7031-8039] {Liton Majumdar}
\affiliation{Exoplanets and Planetary Formation Group, School of Earth and Planetary Sciences, National Institute of Science Education and Research, Jatni 752050, Odisha, India}
\affiliation{Homi Bhabha National Institute, Training School Complex, Anushaktinagar, Mumbai 400094, India}

\begin{abstract}

Over the past three decades, hundreds of rocky exoplanets have been discovered. Some of these show atmospheric signatures indicative of diverse chemical compositions. Interpreting these atmospheres requires a physically grounded understanding of planetary interiors, as interior composition and mineralogy govern the formation and evolution of secondary atmospheres. Rocky terrestrial exoplanets are expected to inherit the refractory composition of their host stars, providing a direct pathway to constrain their bulk composition and mineralogy. However, a large fraction of these planets orbit M-type stars, whose compositions remain poorly constrained because of limited and uncertain stellar abundance measurements. Here, we present a physically consistent framework that connects stellar abundances to the interior structure and mineralogy of rocky exoplanets by combining stellar abundance inference, devolatilization modeling to estimate bulk and mantle elemental abundances, interior structure calculations to derive pressure--temperature profiles, and thermodynamic equilibrium modeling of mantle mineralogy. We first benchmark the framework against Earth and then apply it to the super-Earth GJ 486b using chemically consistent abundances derived from ensembles of similar M-dwarf hosts. We find that GJ 486b likely hosts an iron-rich and silica-poor mantle relative to Earth while preserving the major mantle phase transitions. Sensitivity analyses show that the overall mineralogical structure is robust to variations in bulk composition and pressure--temperature profiles, with temperature primarily modulating phase proportions near key transitions. Our results demonstrate that stellar chemical abundances, when combined with physically motivated devolatilization models, provide robust constraints on planetary interior composition and mineralogy beyond those obtainable from mass and radius measurements alone. This framework establishes a physically motivated foundation for linking rocky exoplanet interiors to atmospheric characterization in the JWST era.

\end{abstract}


\section{Introduction} \label{sec:intro}

More than 6,000 exoplanets have been discovered to date, and this number is expected to grow substantially over the next decade with continued advances in observational capabilities. Over the past two decades, space-based transit surveys and precision radial-velocity programs have yielded large samples of planets with well-determined masses, radii, and orbital properties \citep{Borucki2010_Science_Kepler, Batalha2011_ApJ_Kepler10b, Ricker2015_JATIS_TESS}. These surveys have revealed that small planets, particularly Earth-sized and super-Earth planets, are common in the Galaxy. With the advent of the \textit{James Webb Space Telescope} (JWST), transmission and emission spectroscopy of small exoplanets is becoming feasible, providing the first observational constraints on the atmospheric composition of rocky worlds that possess atmospheres \citep{espinoza2025highlights}. In parallel, mid-infrared instruments on next-generation extremely large telescopes (ELTs) will enable direct imaging and atmospheric characterization of nearby terrestrial planets. Together, these observational advances are shifting the field from the detection and bulk characterization of exoplanets toward probing their atmospheric composition, climate processes, and potential surface environments \citep{Greene2016_ApJ_JWST, Quanz2015_IJAs_METIS}.

Interpreting these atmospheric measurements, however, requires a deeper understanding of the interior structure and geophysical evolution of rocky planets. Terrestrial planets play a central role in studies of planetary formation and evolution because their solid surfaces and silicate-metal interiors can sustain long-lived surface water, active geochemical cycling, and climate feedback (see \citet{baumeister2025fundamentals} for a review). During planetary accretion and early evolution, global magma oceans regulate the storage and release of volatiles, while subsequent volcanism continues to exchange gases between the mantle and the atmosphere. The efficiency and chemical speciation of outgassing depend on several factors, including mantle composition, potential temperature, melt production, redox state, and the rate of volcanic activity \citep{lichtenberg2024super}. Among these, the mineralogical composition of the mantle is particularly important because it governs planetary structure, thermodynamic properties, and volatile inventories, which together influence the composition and evolution of the atmosphere. Therefore, constraining the interior properties of rocky planets is essential for linking atmospheric observations to the underlying geological evolution of these worlds \citep{kite2016atmosphere, VanHoolst2019, nettelmann2021exoplanetary, lichtenberg2025constraining}.

However, constraining the interior properties of rocky planets remains challenging because these quantities cannot be measured directly. Although the masses and radii of many exoplanets are now measured with high precision, their interior structures remain highly degenerate, as different combinations of core size, mantle composition, and layer thicknesses can reproduce the same bulk properties \citep{Rogers2010_ApJ_Degeneracy, dorn2015can}. One promising avenue for breaking this degeneracy is to use the chemical composition of host stars as an additional constraint on the bulk composition of their planets   \citep{Sotin2007, dorn2017generalized, unterborn2019pressure}.

To first order, the bulk composition of a planet is expected to reflect that of its host star, since both form from the same protoplanetary disk \citep{wang2019volatility, liu2020detailed, wang2022detailed}. 
Consequently, stellar abundance measurements provide an important constraint on the expected mineralogy and interior structure of exoplanets. In particular, refractory and moderately refractory elements trace the building blocks of rocky planets, such that stellar abundances of elements including Mg, Si, Fe, Al, Ca, Ni, Ti, Mn, Co, Na, and K provide physically motivated constraints on planetary bulk composition. 
Because stars and their planets originate from the same protoplanetary material, refractory element ratios are expected to be broadly preserved during condensation, with only modest fractionation arising from disk chemistry, migration, and planetary differentiation. \citet{Lodders2003} showed that solar photospheric and CI chondrite abundances agree to within 10\% for 31 of 56 elements and within 15\% for 41 of 56 elements. 
Furthermore, \citet{wang2019enhanced} found that reliable inferences of rocky-planet compositions from stellar refractory abundances require abundance uncertainties below approximately 0.04 dex, corresponding to about 10--15\%. 
These stellar elemental ratios therefore map onto planetary bulk mantle compositions and can constrain mantle mineralogy, such as the relative proportions of olivine, pyroxene, garnet, and oxide phases, thereby informing interior structure and melting behavior, which in turn affects the oxidation state of the mantle \citep{Guimond2023mineralogical, mojzsis2022, unterborn2017effects, spaargaren2023plausible}.
Through their influence on melt composition and volatile solubility, these mineralogical constraints also affect volcanic outgassing 
\citep{hirschmann2012solubility, wordsworth2018redox, ortenzi2020mantle, brugman2021experimental}.

\begin{figure*}[ht!]
    \centering
    \includegraphics[width=\linewidth]{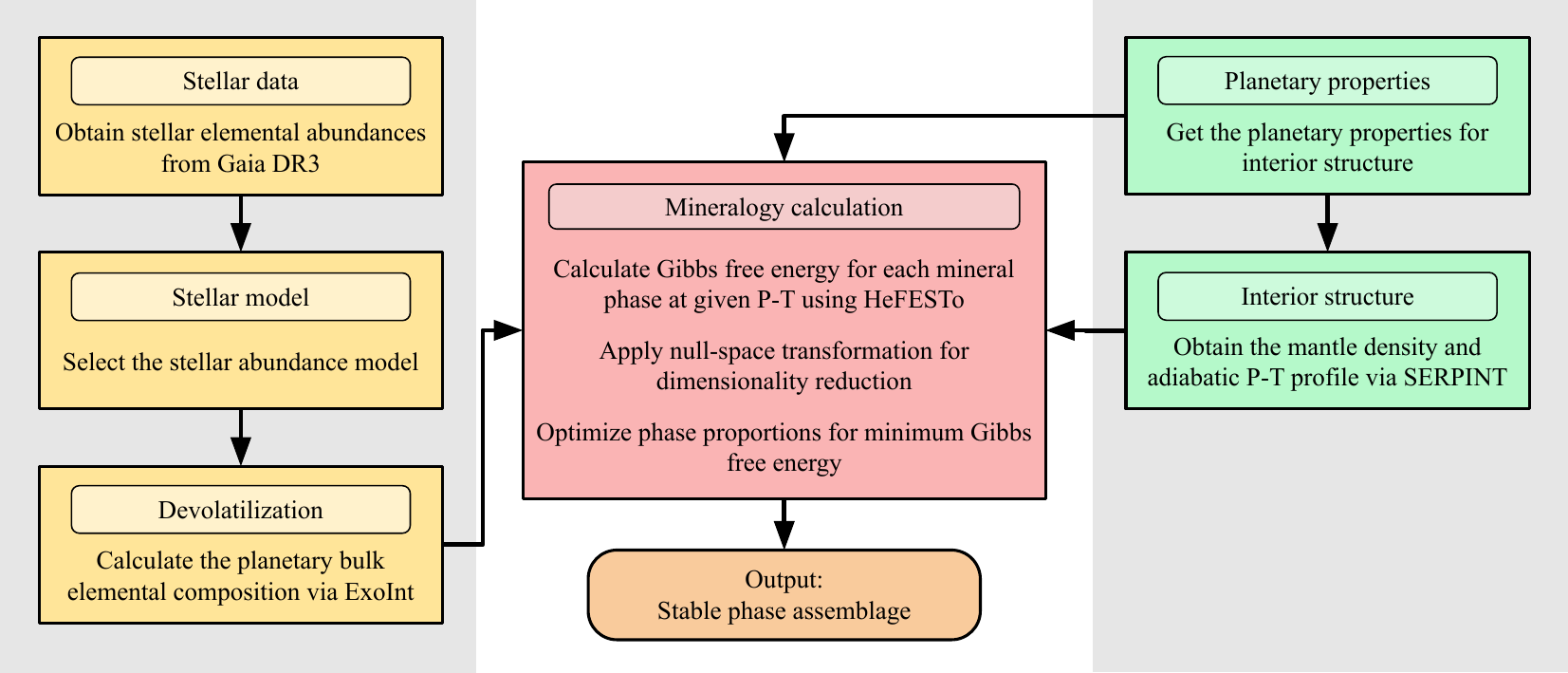}
    \caption{Flowchart of our planetary interior mineralogy framework combining \texttt{ExoInt} \citep{wang2019volatility}, \texttt{SERPINT} \citep{sahu2025unveiling}, and \texttt{HeFESTo} \citep{Stixrude-I, Stixrude-II, Stixrude-III}. This three-stage approach integrates stellar and planetary abundances (left), interior structure (right), and mineralogical calculations (center) to derive the stable phase assemblage of a rocky planet mantle.}
    \label{fig:flowchart}
\end{figure*}

M-dwarf stars are particularly important in this context because a large fraction of the rocky planets discovered to date orbit such low-mass stars. Their small sizes, low luminosities, and the short orbital periods of their planets produce deep transits and large radial-velocity signals, thereby enabling detailed observational characterization of rocky worlds. Moreover, M dwarfs dominate the local stellar population, comprising roughly $70\%$ of stars in the Milky Way and the solar neighborhood according to volume-limited surveys and field luminosity functions \citep{henry2024character}. For these stars, the \textit{Gaia} Data Release 3 (DR3) catalog provides high-precision astrometry, broadband photometry, and homogeneous astrophysical parameters for hundreds of millions of stars within the solar neighborhood ($\sim150$ pc). The catalog includes derived stellar properties such as effective temperature, luminosity, radius, and mass and, for a significant subset, also provides estimates of bulk metallicity and individual elemental abundances obtained through spectrophotometric modeling \citep{vallenari2023gaia, behmard2025data}.

Motivated by these recent observational advances, our goal is to develop a physically consistent framework that connects host-star abundances to the interior structure and mineralogy of rocky exoplanets. To achieve this, we combine stellar abundance inference, devolatilization modeling, interior structure calculations, and thermodynamic equilibrium mineralogy. Specifically, we use stellar abundance data from the \textit{Gaia} DR3 catalog \citep{vallenari2023gaia,behmard2025data} and the Hypatia Catalog \citep{Hinkel2014} to estimate the bulk and mantle elemental abundances of rocky exoplanets using the devolatilization model \texttt{ExoInt} \citep{wang2019volatility, wang2019enhanced}. We then compute the pressure--temperature ($P$--$T$) profile using the rocky planet interior structure and evolution model \texttt{SERPINT} \citep{sahu2025unveiling}. Finally, we combine these results with \texttt{HeFESTo} \citep{Stixrude-I, Stixrude-II, Stixrude-III}, a Gibbs free-energy minimization framework for calculating thermodynamic equilibrium in multi-component mineral systems, to predict the mantle mineralogy as a function of planetary depth.

We apply this integrated \texttt{ExoInt}--\texttt{SERPINT}--\texttt{HeFESTo} framework (hereafter the ESH framework) to GJ 486b, a potentially rocky planet with a mass of $3.00\,M_{\oplus}$ and a radius of $1.343\,R_{\oplus}$, orbiting the M3.5V star GJ 486, which has a radius of $0.339\,R_{\odot}$ and an effective temperature of 3317 K \citep{caballero2022detailed}. Using this framework, we derive plausible mantle mineralogy profiles as a function of planetary depth and perform sensitivity analyses to assess the robustness of the predicted mineralogy to uncertainties in both bulk composition and pressure--temperature structure.

Estimating the mantle mineralogy of rocky exoplanets directly from host-star abundances presents several methodological challenges, two of which are addressed in this work. First, precise elemental abundance measurements for M-dwarf stars remain difficult to obtain, and even modest observational uncertainties can propagate into significant uncertainties in the inferred mantle oxide composition and resulting mineralogy. To address this challenge, we develop a robust statistical framework for inferring stellar elemental abundances and quantifying how these uncertainties propagate into the bulk and mantle compositions of their orbiting rocky planets. Second, we perform a comprehensive sensitivity analysis to assess the robustness of the predicted mantle mineralogy to uncertainties in both bulk composition and pressure--temperature structure. Collectively, these developments provide a practical and physically consistent framework for estimating the mantle mineralogy of rocky exoplanets from fundamental stellar and planetary properties, establishing a foundation for connecting host-star abundances to planetary interiors and, ultimately, to their atmospheric characterization in the JWST era.

The structure of this paper is as follows. In Section \ref{sec:methods}, we describe the methodology used to infer mantle mineralogy from stellar abundance data. Section \ref{sec:results} presents the principal results and sensitivity analyses. In Section \ref{sec:discussion}, we discuss the implications and limitations of the framework. Finally, Section \ref{sec:conclusion} summarizes the main conclusions of this work.

\begin{figure*}[ht!] 
    \centering
    \includegraphics[width=\linewidth]{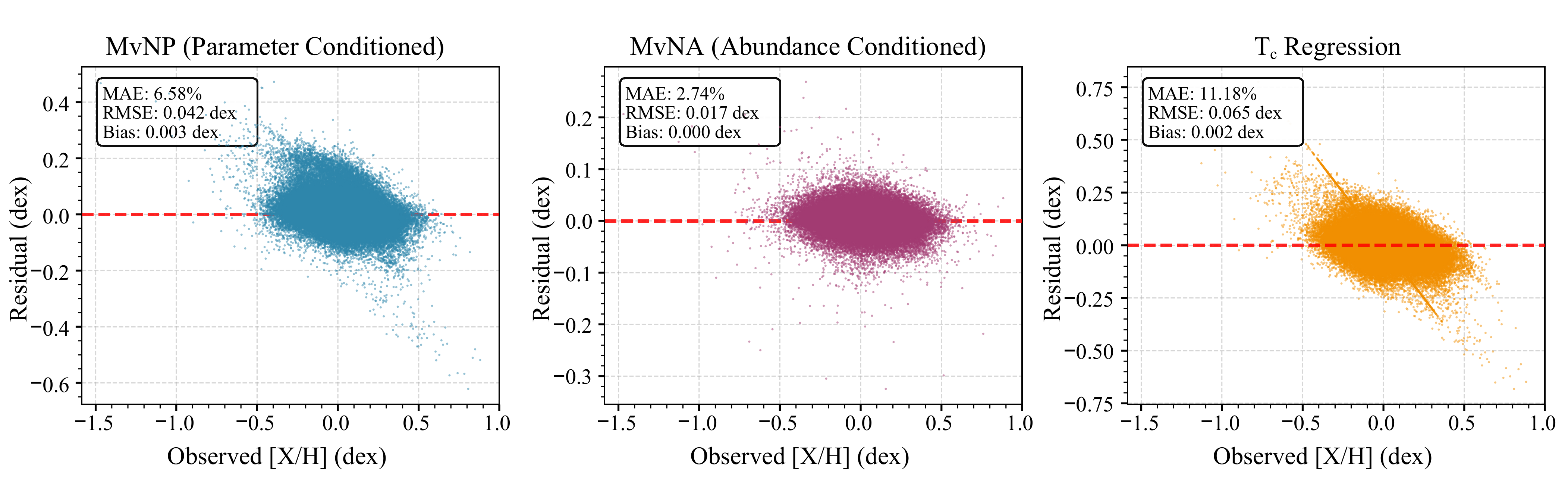} 
    
    \caption{Residual (MAE: mean absolute error) distributions for the stellar abundance prediction models. Residuals are defined as $\Delta[\mathrm{X/H}] = [\mathrm{X/H}]_{\mathrm{pred}} - [\mathrm{X/H}]_{\mathrm{obs}}$.
The dashed horizontal line marks zero residual. The MvNP and MvNA models show the tightest residual distributions and the lowest scatter. The tight and unbiased residual distribution of the MvNA model indicates that element--element correlations provide stronger predictive power than parameter-based trends alone, while the $T_c$ regression shows larger dispersion and systematic errors.}
\label{fig:residual}
\end{figure*}

\section{Methods} \label{sec:methods}
We present a computational framework that uses stellar chemical abundances to predict the mineralogy of rocky exoplanet mantles. The model proceeds in three main stages: (1) calculating the planetary elemental composition from bulk stellar abundances, (2) constructing the interior structure of the target planet, and (3) determining the mantle mineralogy under equilibrium conditions. Figure \ref{fig:flowchart} provides an overview of the computational framework. We describe each stage in detail in the subsections below.

\subsection{Stellar Abundance Data} \label{subsec:stellar_abundance_data}

In this work, stellar chemical abundances are used to define the bulk composition of rocky exoplanets. We focus primarily on refractory elements with high 50\% condensation temperatures ($T_c^{50} \gtrsim 1300$~K; \citealt{Lodders2003}), along with a few moderately volatile elements. In total, we consider Mg, Si, Fe, Ca, Al, Na, Cr, O, C, N, Ni, and Ti, which together dominate the mass budget of terrestrial mantles and cores \citep{Bond2010, Thiabaud2015, unterborn2016scaling}. All twelve elements are included in the stellar abundance model described in Section~\ref{subsec:stellar_abundance_models}. Of these, only Mg, Si, Fe, Ca, Al, Na, Cr, and O are subsequently passed to \texttt{HeFESTo} to compute the mantle oxide composition.
Mantle minerals bearing the remaining four elements (Ni, Ti, N, and C) are not included in the \texttt{HeFESTo} database \citep{Stixrude-III}. Elemental abundance ratios such as Mg/Si and Fe/Mg are computed from these data and serve as key inputs for constraining mantle mineralogy, phase proportions, and core-mantle differentiation \citep{Sotin2007, unterborn2016scaling, unterborn2019pressure, dorn2017generalized, Guimond2023mineralogical, guimond2024stars, spaargaren2023plausible, putirka2019composition}. 
These ratios are propagated through the interior and the mineralogical models described in subsequent sections.

We adopt stellar abundance measurements from the data-driven analysis of M dwarfs presented by \citet{behmard2025data} alongside compiled literature catalogs from \citet{Hinkel2014}, which provides detailed elemental abundances for approximately $17,000$ stars observed as part of the Sloan Digital Sky Survey (SDSS-V). 
This dataset combines high-resolution near-infrared spectroscopy with \textit{Gaia} DR3 stellar parameters and a data-driven spectral modeling framework, enabling abundance determinations across a wide range of effective temperatures and metallicities. 

From this catalog, we extract logarithmic elemental abundances relative to solar values (e.g., [Fe/H], [Mg/H], [Si/H], along with other refractory and moderately volatile elements where available). We restrict our sample to stars with reliable effective temperatures and metallicities.

To account for missing values and heterogeneous uncertainties across elements, we employ a set of complementary abundance prediction models to estimate incomplete measurements in a statistically consistent manner.

The final set of stellar abundances is used to compute elemental ratios and define bulk refractory compositions, which are then used as inputs for the planetary interior structure and equilibrium mineralogy calculations.

\subsection{Stellar Abundance Models} \label{subsec:stellar_abundance_models}

Elemental abundance measurements for M-dwarf stars are often incomplete, with many elements either unmeasured or subject to large uncertainties. To obtain a chemically consistent set of stellar abundances suitable for planetary interior modeling, we employ a suite of complementary statistical models to estimate missing elemental abundances: (i) a multivariate normal (Gaussian) model conditioned on stellar parameters (MvNP), (ii) a multivariate conditional abundance model (MvNA), and (iii) a condensation-temperature-based regression model ($T_c$ regression). All models are applied directly to the M-dwarf abundance dataset described in Section \ref{subsec:stellar_abundance_data}.

For the multivariate normal (MvN) models (i.e., both MvNP and MvNA), which rely on empirical abundance correlations, we define a ``local neighbor'' sample for each target star by selecting stars within $|\Delta T_{\mathrm{eff}}| \le 100$ K and $|\Delta[\mathrm{Fe/H}]| \le 0.05$ dex. This criterion ensures that abundance correlations are estimated from stars occupying a similar region of stellar parameter space. Figure \ref{fig:residual} shows the distribution of residuals for all stars included in these models.

\begin{figure*}[ht!] 
    \centering
    \includegraphics[width=\linewidth]{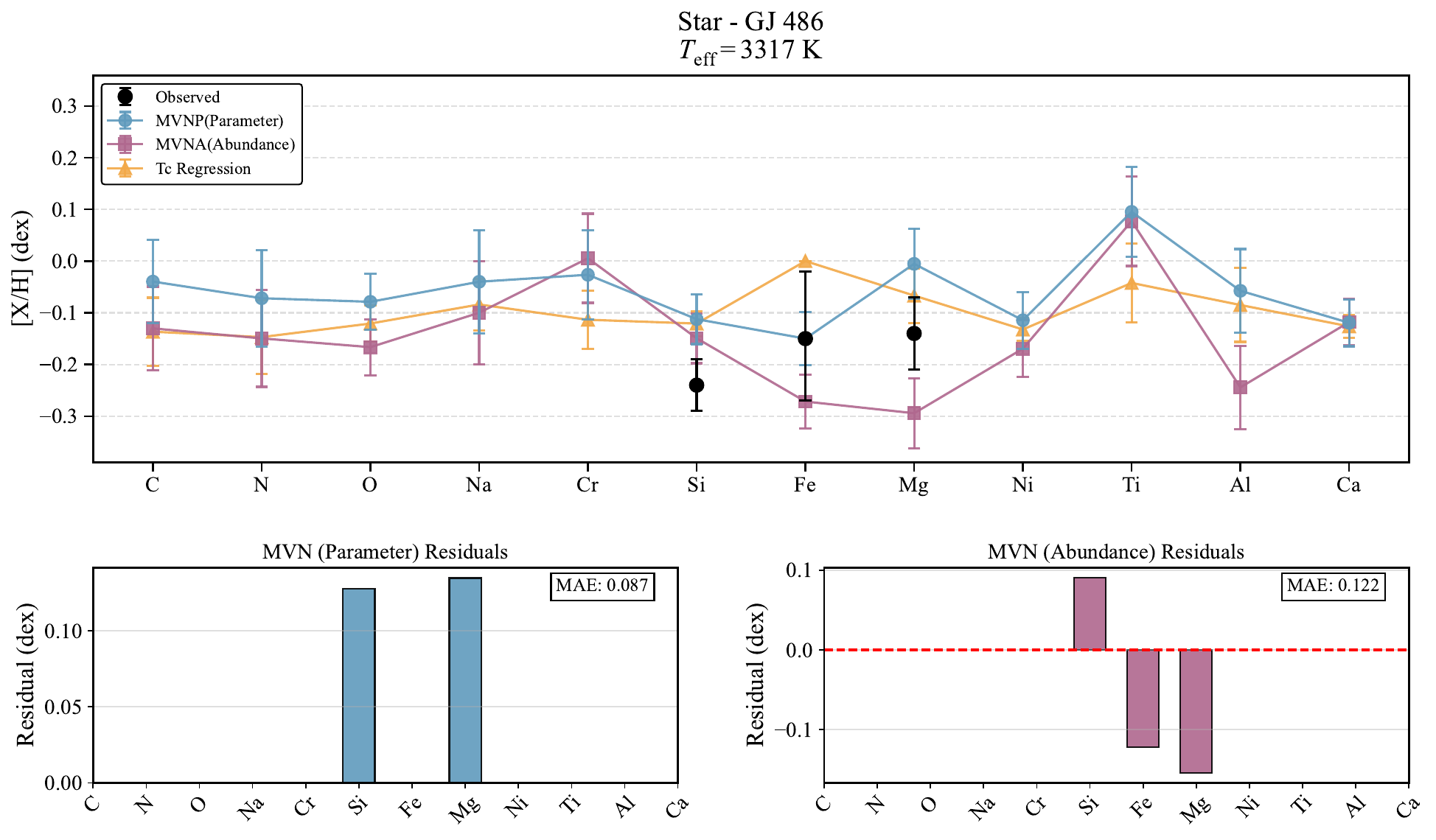}
 \caption{Predicted elemental abundances for the M-dwarf GJ 486 ($T_{\mathrm{eff}} = 3317\,\mathrm{K}$) compared with observed values. 
The top panel shows logarithmic abundances [$\mathrm{X/H}$] predicted by three models: a parameter-conditioned multivariate normal model (MvNP; blue circles), an abundance-conditioned multivariate normal model (MvNA; magenta squares), and a condensation-temperature regression model (orange triangles). 
Observed abundances are shown as black circles where available. 
The bottom panels show the residuals (predicted minus observed abundances) for the MvNP (left) and MvNA (right) models, with the dashed red line indicating zero residual. 
Mean absolute errors (MAE), computed over elements with available observations, are indicated in each panel. All abundances are reported relative to solar values in logarithmic (dex) units.}
    \label{fig:Predicted_elemental_abundances}
\end{figure*}

\subsubsection{Conditioned Multivariate Normal Models (MvNP and MvNA)} \label{subsubsec:multivariate}

Elemental abundances in stars are correlated due to common nucleosynthetic origins and chemical evolution, and they vary systematically with stellar metallicity and effective temperature. The MvNP model exploits these correlations by modeling the joint distribution of abundances and stellar parameters.

We distinguish between two implementations of the multivariate normal framework based on the information used for conditioning. In the parameter-conditioned model (MvNP), elemental abundances are predicted using only the observed stellar parameters, $\boldsymbol{\theta} = ([\mathrm{Fe/H}], T_{\mathrm{eff}})$, and the empirical abundance-parameter correlations inferred from the local neighbor sample. This approach is applicable when only stellar parameters are available or when abundance measurements are incomplete for the target star. It provides a baseline prediction driven by global chemical trends in the dataset.

In contrast, the abundance-conditioned model (MvNA) conditions the multivariate distribution on a subset of observed elemental abundances for the target star, rather than on stellar parameters alone. 
By exploiting intrinsic element--element correlations arising from common nucleosynthetic origins and Galactic chemical evolution, MvNA leverages partial abundance information (e.g., Fe, Mg, or Si) to predict the remaining elements. 
As a result, MvNA typically yields improved accuracy when even a small number of elemental abundances are available, while MvNP remains essential for stars with only basic stellar parameters.

Let $\mathbf{X} = (X_1, \ldots, X_p)$ denote the vector of logarithmic elemental abundances relative to solar, where $X_j \equiv [\mathrm{x}_j/\mathrm{H}]$, and let $\boldsymbol{\theta} = ([\mathrm{Fe/H}], T_{\mathrm{eff}})$ denote the observed stellar parameters. 
We assume that abundances and stellar parameters follow a joint multivariate normal distribution. Logarithmic abundance ratios in stellar populations are often approximately Gaussian distributed after normalization to solar values \citep{casey2019data, Ness2015}.

\[
\begin{pmatrix}
\mathbf{X} \\
\boldsymbol{\theta}
\end{pmatrix}
\sim
\mathcal{N}
\left(
\begin{pmatrix}
\boldsymbol{\mu}_X \\
\boldsymbol{\mu}_{\theta}
\end{pmatrix},
\begin{pmatrix}
\Sigma_{XX} & \Sigma_{X\theta} \\
\Sigma_{\theta X} & \Sigma_{\theta\theta}
\end{pmatrix}
\right),
\]

where $\boldsymbol{\mu}_X$ and $\boldsymbol{\mu}_{\theta}$ are the mean abundances and mean stellar parameters, respectively, estimated from the local neighbor sample. 
The covariance blocks $\Sigma_{XX}$, $\Sigma_{\theta\theta}$, $\Sigma_{\theta X}$, and $\Sigma_{X\theta}$ encode element--element correlations, parameter dispersion, and abundance--parameter correlations.

For a target star with observed parameters $\boldsymbol{\theta}_\star$, the conditional mean abundance vector is
\[
\mathbb{E}[\mathbf{X} \mid \boldsymbol{\theta}=\boldsymbol{\theta}_\star]
=
\boldsymbol{\mu}_X
+
\Sigma_{X\theta}\Sigma_{\theta\theta}^{-1}
(\boldsymbol{\theta}_\star - \boldsymbol{\mu}_{\theta}),
\]
which shifts the local mean abundance pattern according to the deviation of the target star's parameters from the mean of the neighbor sample.

\subsubsection{Condensation temperature regression}
The condensation temperature ($T_c$) model provides a physically motivated ordering of stellar elements according to their volatility during disk cooling, separating volatile species from refractory solids. Trends with condensation temperature have been proposed as potential signatures of dust and gas fractionation or planet formation processes. Accordingly, trends with $T_c$ are expected to emerge. The $T_c$ regression model therefore serves as a simple and interpretable baseline for testing whether elemental abundances follow condensation-temperature-driven behavior.

For each element, we fit a linear relation
\[
[\mathrm{X/H}] = a + b\,T_c + c\,[\mathrm{Fe/H}],
\]
where $T_c$ is the 50\% condensation temperature in a solar-composition gas \citep{Lodders2003}. The coefficient $b$ captures any systematic dependence on condensation temperature, while $c$ accounts for overall metallicity scaling. The model coefficients are estimated independently for each element using ordinary least squares applied to all stars with available measurements.

\subsubsection{Model Selection} \label{subsubsec:model_selection}
Across the population, the MvNA model achieves the lowest mean absolute error (MAE) for most elements (see Figure \ref{fig:residual}), particularly for refractory species (e.g., Si, Fe, Mg, Ca, and Ti). 
This suggests that abundance correlations remain useful for predicting planetary composition when only a subset of elemental abundances is available. Typical errors for MvNA are $\sim$0.02--0.05 dex, with minimal bias. The MvNP model performs consistently across all elements, with moderate MAEs of $\sim$0.05--0.10 dex, reflecting its reliance on smooth abundance trends with $T_{\mathrm{eff}}$ and [Fe/H], and yielding errors comparable to those of the MvNA model.

In contrast, the $T_c$ regression model shows substantially larger errors, particularly for iron and other refractory elements, indicating that condensation temperature alone cannot accurately capture the detailed chemical structure of M-dwarf abundances (as shown in Figure \ref{fig:residual}).

 Figure \ref{fig:Predicted_elemental_abundances} shows an example prediction for the M-dwarf GJ~486. For this individual case, the parameter-conditioned model (MvNP) produces a slightly lower MAE than the abundance-conditioned model (MvNA). This partly arises because \text{[Fe/H]} is explicitly used as an input parameter in MvNP, effectively constraining the predicted Fe abundance and reducing its contribution to the average error. In contrast, MvNA treats Fe as a component of the abundance vector within the multivariate correlation structure, allowing the predicted value to deviate from the observed abundance. Specifically, in this application of MvNA, \text{[Si/H]} is predicted by conditioning on the observed \text{[Mg/H]} and \text{[Fe/H]}, \text{[Mg/H]} by conditioning on the observed \text{[Si/H]} and \text{[Fe/H]}, and \text{[Fe/H]} by conditioning on the observed \text{[Si/H]} and \text{[Mg/H]}. The resulting non-zero residuals therefore reflect the degree to which the specific elemental abundance ratios of GJ~486 (e.g., Si/Mg) deviate from the conditional mean inferred from chemically similar M-dwarfs with comparable $T_{\mathrm{eff}}$ and \text{[Fe/H]}. 

The MvNA model nevertheless produces the most compact and symmetric residual distribution, while the MvNP model remains largely unbiased but shows a broader dispersion. This behavior reflects the different information used by the two models: MvNP relies primarily on global abundance trends with stellar parameters ($T_{\mathrm{eff}}$ and [Fe/H]), whereas MvNA additionally exploits intrinsic element--element correlations present in the stellar abundance distribution of the training sample. The local neighbor sample used to estimate these correlations is defined in stellar parameter space using Gaia DR3 values of $T_{\mathrm{eff}}$ and [Fe/H].

Rather than selecting a single preferred model for systems with incomplete abundance measurements, both approaches can be used to explore the range of plausible stellar compositions. In such cases, MvNP and MvNA together provide complementary estimates that span the plausible chemical regimes of the host star and, consequently, the range of mantle compositions that may arise in the corresponding rocky exoplanets.

The $T_c$ regression model displays both increased scatter and systematic residual trends, underscoring its role as a physically interpretable baseline rather than a high-accuracy predictor. For this reason, we do not use the $T_c$ regression model for mineralogy calculations in this study.

\subsection{Inferring Planetary Bulk Composition} \label{subsec:inferring_bulk_composition}
 
Stellar elemental abundances represent the initial chemical reservoir of a protoplanetary disk, but they cannot be used directly as a proxy for planetary compositions. During disk evolution, elements condense at different temperatures, and solid material forming at different locations undergoes varying degrees of thermal processing. This leads to systematic chemical fractionation before planet assembly, causing rocky planets to accrete material that differs in composition from their host stars \citep{Bond2010, Thiabaud2015, wang2019volatility}. A primary driver of this fractionation is devolatilization, in which elements with lower condensation temperatures are progressively depleted relative to refractory elements. This process is controlled by the disk temperature structure and formation location, rather than by the later stages of planet growth \citep{Lodders2003, wang2019enhanced, suer2023distribution}. Devolatilization therefore sets the bulk chemical makeup of rocky planets.

We model this transformation using the open source code \texttt{ExoInt} \citep{wang2019enhanced, wang2022detailed}, which applies volatility-dependent depletion and subsequently accounts for metal-silicate differentiation. The model quantifies elemental depletion as a function of the 50\% condensation temperature and is calibrated using the compositional differences between the proto-Sun and the Earth. It thus yields both a devolatilized bulk planetary composition and a corresponding bulk mantle composition. 

In this model, the partitioning of iron between the silicate mantle and 
the metallic core is determined through an oxygen mass-balance calculation, which provides a first-order estimate of metal-silicate differentiation \citep{wang2019enhanced, wang2022detailed}.
In \texttt{ExoInt}, oxygen is first allocated to the major lithophile elements according to their stoichiometric requirements to form silicate and oxide phases. 
The remaining oxygen budget then determines the fraction of iron that is oxidized to FeO in the mantle. Any iron that cannot be oxidized is assigned to the metallic reservoir and contributes to the core. In this framework, the core mass fraction emerges naturally from elemental stoichiometry and oxygen availability, rather than being prescribed a priori. We note that this approach does not explicitly account for processes such as planetary migration, giant impacts, volatile delivery, disequilibrium accretion, or redox evolution during core formation, all of which can influence the Fe/FeO partitioning and the resulting core mass fraction. Consequently, the inferred core mass fraction should be interpreted as a first-order, model-derived estimate rather than a unique prediction of the planet's differentiation history. 
These derived mantle-core compositions provide physically consistent inputs for interior and mineralogical modeling. Elemental ratios such as Fe/Mg and Mg/Si emerge naturally from the devolatilized composition. The Fe/Mg ratio is determined by the resulting partitioning of iron between the core and mantle, while the Mg/Si ratio controls the dominant silicate mineralogy. 

\begin{figure*}[t]
    \centering
    \includegraphics[width=\textwidth]{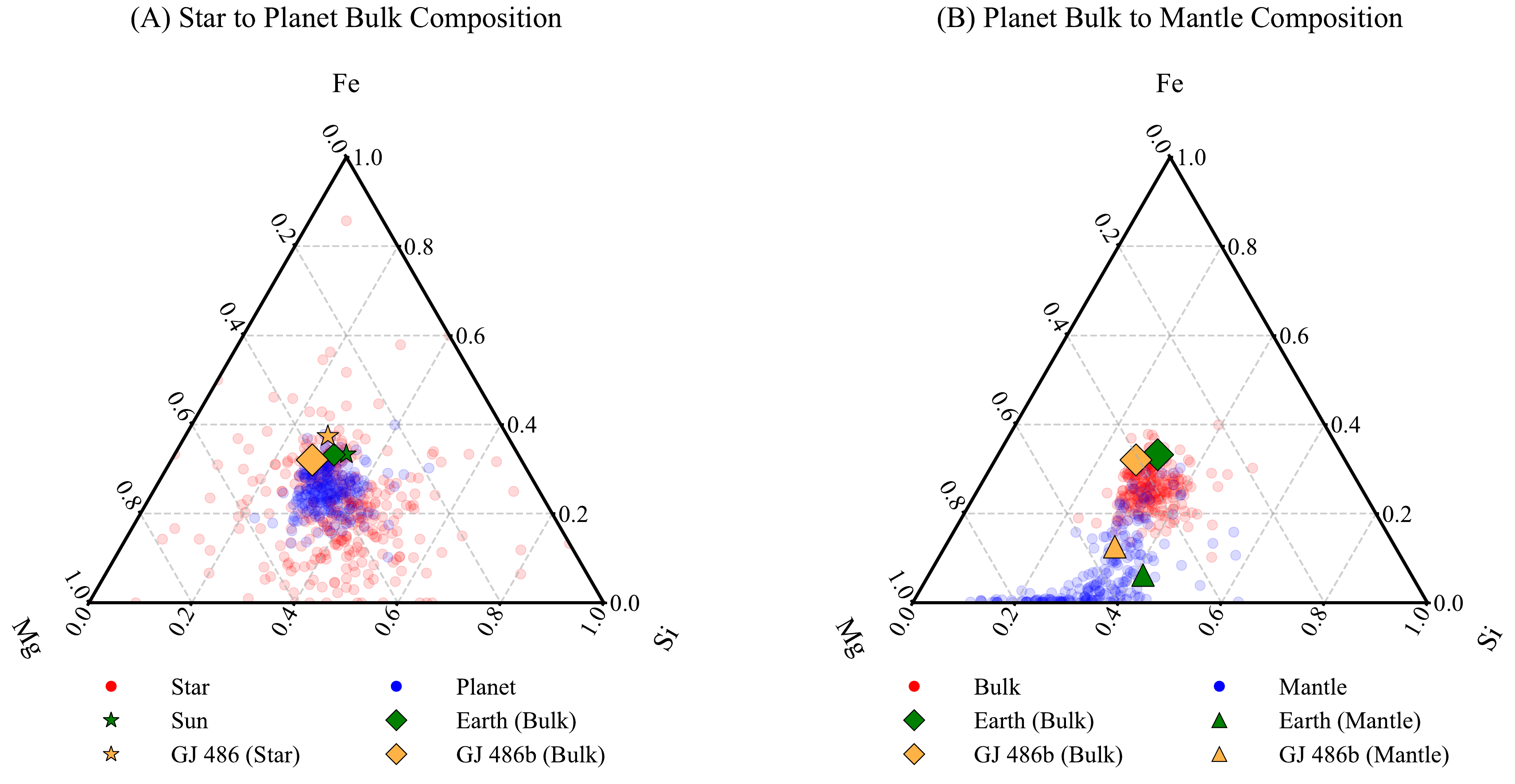}
    \caption{Modeled Fe-Mg-Si molar abundances for the stellar and planetary sample. The left panel contrasts host star metallicity with the modeled bulk composition of orbiting planets, illustrating the correlation between stellar photospheric abundances and planetary bulk refractory ratios. The right panel demonstrates the compositional shift driven by differentiation by comparing initial planetary bulk compositions with their resulting silicate mantles, where iron is sequestered into the core. The Sun-Earth and GJ 486 systems are highlighted as Solar System and exoplanetary reference baselines. Data are derived from the \textit{Gaia} DR3 catalog.}
    \label{fig:ternary_plots}
\end{figure*}

The latest publicly available version of \texttt{ExoInt} does not include chromium in its devolatilization pipeline by default, as Cr$_2$O$_3$ is a relatively low-abundance oxide compared to the major mantle components.
As Cr is nearest to Fe and Si in terms of 50\% condensation temperature ($T_{\mathrm{c},50}$), to incorporate Cr, we extrapolated its devolatilization coefficients from those of Si and Fe, following the same methodology on which \texttt{ExoInt} is built \citep{wang2022detailed}. We find that omitting Cr$_2$O$_3$ from the bulk mantle composition alters the oxide stoichiometry sufficiently to stabilize free metallic iron in the \texttt{HeFESTo} Gibbs-energy minimization. This arises from the more reduced bulk metal-to-oxygen ratio resulting from the removal of the trivalent cation Cr$^{3+}$. A similar effect on Fe-metal stability can arise from variations in the abundance of the trivalent oxide Fe$_2$O$_3$ \citep{dong2026metal}. However, systematically exploring variations in Cr$^{3+}$ and Fe$^{3+}$ concentrations and their effects on Fe-metal stability is beyond the scope of this study. While incorporating Cr$_2$O$_3$ prevents the formation of a free metallic iron phase for the specific bulk composition considered here, we note that free Fe-metal may also occur in physically plausible mantle mineralogies under sufficiently reducing conditions. 

Figure \ref{fig:ternary_plots} shows the Fe--Mg--Si ternary diagram comparing stellar, planetary bulk, and planetary mantle compositions for systems within the Gaia DR3 catalog. Figure \ref{fig:ternary_plots}A compares stellar compositions with the bulk compositions of their planets, while Figure \ref{fig:ternary_plots}B compares planetary bulk compositions with the corresponding average mantle compositions.

\subsection{Setting-up the Interior Structure} \label{subsec:setting_interior}

We use a time snapshot of the $P$--$T$ profile of a solidified planet obtained from the \texttt{SERPINT} code \citep{sahu2025unveiling}. 
\texttt{SERPINT} is a one-dimensional, self-consistent model of rocky planet structure and thermal evolution. 
Given a planet's physical and thermodynamic properties, it solves the coupled equations of hydrostatic equilibrium, mass conservation, and energy transport to determine the interior structure, core size, and thermal evolution from the magma ocean stage to a fully solidified state.

We obtain the planet's internal structure, density profile as a function of its depth, and the corresponding pressure-temperature (P-T) profile at the end of the \texttt{SERPINT} model run. 
This adiabatic $P$--$T$ profile is representative of the bulk mantle and increases linearly with depth, as described in Equation 12 of \citet{sahu2025unveiling} and Figure \ref{fig:PT_planets}.
This is then used as input to \texttt{HeFESTo} to compute the depth-dependent mineral phase assemblages. 

However, \texttt{SERPINT} assumes an Earth-like bulk composition when computing the interior structure. 
This assumption may not always hold for rocky planets orbiting different types of host stars. Previous studies, such as \citet{seager2007mass, unterborn2016scaling} show that variation in a planet's bulk composition can lead to changes in its interior structure and, consequently, its density and pressure profiles. Nevertheless, the resulting pressure-temperature (P-T) structure has been shown to be relatively insensitive to moderate variations in bulk composition and interior structure \citep{unterborn2019pressure, baumeister2025fundamentals}. This justifies the use of the computed adiabatic $P$-$T$ profile in this work.

\subsection{Deducing the Mantle Mineralogy using \texttt{HeFESTo}}
\label{subsec:hefesto}

To compute the stable mineralogical assemblage of the mantle under prescribed $P$–$T$ conditions, we employ \texttt{HeFESTo} (Helmholtz Free Energy Self-consistent Thermodynamics), a Gibbs free energy minimization framework developed to model the thermodynamic equilibrium of multicomponent mineral systems \citep{Stixrude-I, Stixrude-II, Stixrude-III}. 
The algorithm determines the equilibrium phase assemblage by minimizing the total Gibbs free energy of the system, subject to constraints imposed by bulk composition, pressure, and temperature.
At each point along the planetary interior $P$--$T$ profile, \texttt{HeFESTo} evaluates all permissible phase combinations and identifies the assemblage that satisfies thermodynamic equilibrium while conserving mass \citep{Stixrude-II}. 

The theoretical basis of this approach follows classical equilibrium thermodynamics, in which the stable state of a system at fixed $P$--$T$ conditions corresponds to a minimum in Gibbs free energy.
This formulation enables pressure- and temperature-dependent variations in mineral stability to be treated consistently throughout the interior.
Individual mineral phases within \texttt{HeFESTo} are represented as solid solutions composed of multiple endmember species, corresponding to chemically pure compositional endmembers of each phase. 
The chemical potential of each solid-solution species includes contributions from the Gibbs free energy of the pure end-member, ideal entropy, and non-ideal excess interactions, following the solution model implemented in \texttt{HeFESTo} \citep{Stixrude-II}. This treatment allows continuous compositional variability within mineral phases and avoids artificial discretization of phase boundaries. Similar thermodynamic approaches have been implemented in codes such as \texttt{Perple\_X} \citep{connolly1990multivariable, connolly2005computation}, \texttt{MELTS} \citep{ghiorso1995chemical, ghiorso1997thermodynamic}, and \texttt{pMELTS} \citep{ghiorso2002pmelts, balta2013application}. More broadly, the overall framework adopted in this study follows a two-step approach: translating host-star elemental abundances into exoplanet mantle oxide stoichiometry, and subsequently using these compositions as inputs to Gibbs free-energy minimization models, such as \texttt{HeFESTo} or \texttt{Perple\_X}, to estimate mantle phase abundances and structure. This framework builds directly upon established methodologies from earlier studies \citep{dorn2015can, dorn2017generalized, unterborn2017effects, spaargaren2020influence, spaargaren2023plausible, Guimond2023mineralogical, wang2022detailed}.

\begin{figure*}[ht!]
    \centering
    \includegraphics[width=\linewidth]{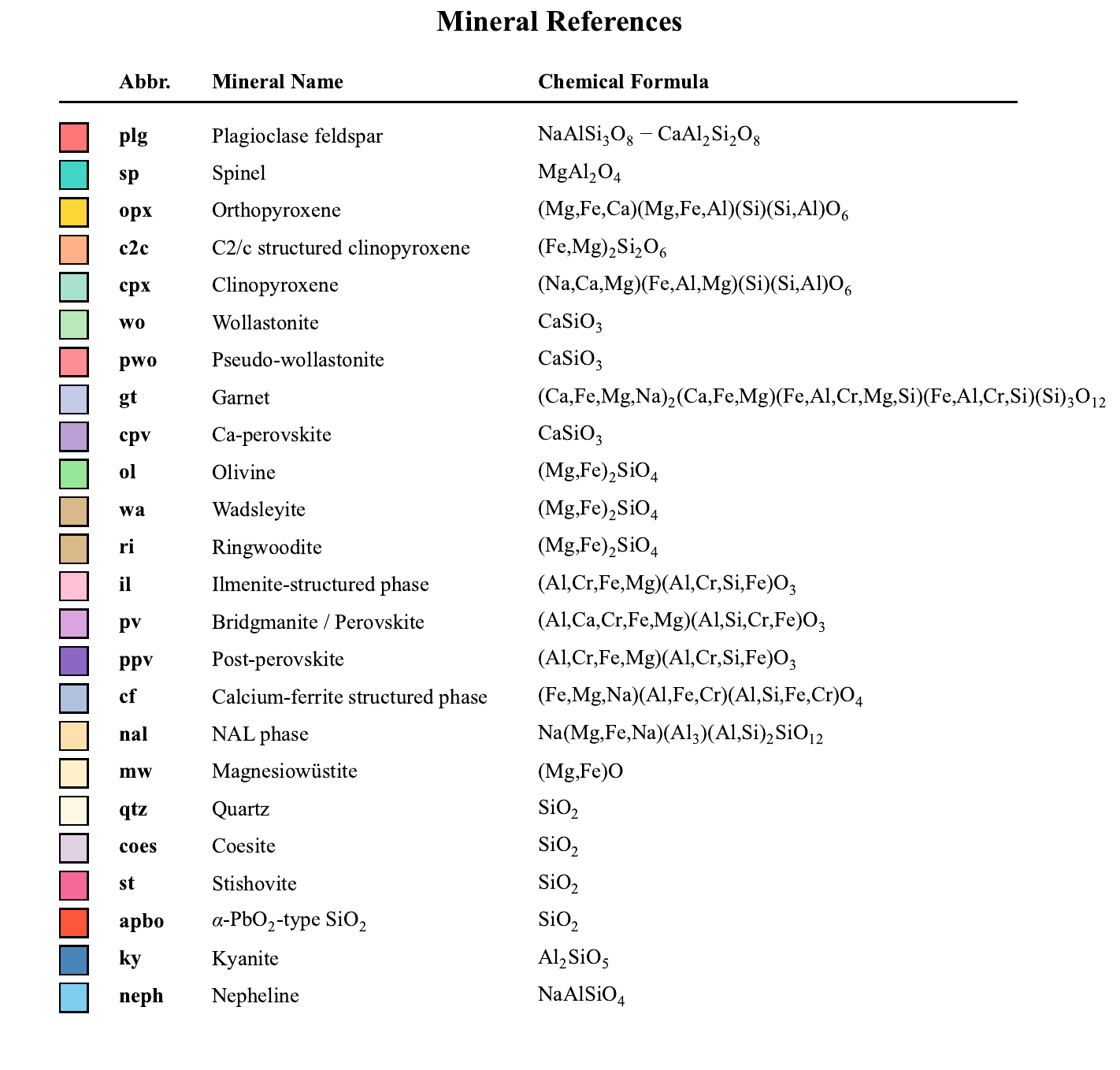}
    \caption{Reference mineral phases used in this study, showing their abbreviations, full mineral names, and representative chemical formulae. The listed phases correspond to the subset of mineral assemblages common to both the devolatilized Earth reference composition and the GJ 486b bulk compositions explored in this paper. Only phases that attain non-negligible stability over the pressure-temperature range sampled in our models are shown, and these phases are consistently illustrated in the phase diagrams throughout the paper.}
    \label{fig:list_of_minerals}
\end{figure*}

The \texttt{HeFESTo} thermodynamic database contains a broad range of minerals relevant to planetary interiors, with equations of state and thermodynamic parameters calibrated over wide pressure and temperature ranges \citep{Stixrude-I, Stixrude-III}. Recent updates to the database incorporate improved descriptions of iron-bearing phases, including the effects of iron partitioning and high-pressure spin transitions, thereby extending the applicability of the model to non-Earth-like bulk compositions \citep{Stixrude-III}.

In this study, we adopt a physically motivated subset of mineral phases from the (2024) version of the \texttt{HeFESTo} database that is appropriate for rocky planetary mantles \citep{Stixrude-III}. 

This selection is designed to capture the dominant mineralogical behavior relevant to terrestrial and super-Earth interiors while avoiding unnecessary overparameterization of the inferred compositions. The mineral phases considered in this work are shown in Figure \ref{fig:list_of_minerals}.

\subsection{Sensitivity Analysis}
\label{subsec:sensitivity_analysis_method}

While the described framework links stellar chemistry to mantle mineralogy, the predicted phase assemblages depend strongly on both the bulk elemental composition of the host star and the mantle pressure--temperature ($P$--$T$) profile. The $P$--$T$ profile is primarily controlled by mantle potential temperature, convective efficiency, thermal gradients, and the internal density and gravity structure set by the planetary mass and radius.

Because mineral phase stability is determined by Gibbs free energy minimization at given $P$ and $T$, even moderate variations in the mantle $P$--$T$ profile can shift phase boundaries and significantly alter phase proportions (wt\%). It is therefore important to quantify the sensitivity of the predicted mineralogy to plausible uncertainties in the $P$--$T$ profile.
A similar dependence is expected for the bulk elemental composition of GJ 486b. As described in Section \ref{subsec:stellar_abundance_models}, the inferred bulk elemental abundances depend on the adopted model and its associated uncertainties. The calculated oxide compositions are also expected to contain errors inherently. Variability in oxide composition within these errors ranges can directly affect the resulting phase diagrams. 
The upper and lower error bounds might show distinct phases that did not appear in the modal abundances. 
It is thus important to perform a sensitivity analysis of these bounds on the phase diagrams. 

In this work, we perform a controlled $P$--$T$ profile sensitivity analysis by perturbing the baseline temperature profile while keeping the bulk composition fixed. We also conduct a bulk composition sensitivity analysis by evaluating variations within the corresponding compositional uncertainty bounds. This approach allows us to isolate the thermodynamic response of mantle mineralogy to thermal and compositional variations.

\section{Results} \label{sec:results}

We present results for the potentially rocky planet GJ 486b as a case study. In this section, we first discuss the model selection for the host star GJ 486 and then proceed to the inferred bulk composition and corresponding mineralogy of GJ 486b.

\subsection{Stellar Model Selection for GJ 486} \label{subsec:stellar_model_GJ}

In the case of GJ~486b, the \texttt{ExoInt} results show a significant difference between the MvNP and MvNA models (see Figure~\ref{fig:mvna_mvnp_final_gj} and Table~\ref{tab:bulk_composition}). In terms of bulk oxygen availability,
\texttt{ExoInt} \citep{wang2019enhanced} samples the partitioning of oxygen among Fe, Ni, and S using a Monte Carlo approach. The residual unoxidised metal ($\sim 0.511$~wt\% for MvNA; $\sim 0.035$~wt\% for MvNP) and the remaining oxygen after oxide formation ($\sim 0.128$~wt\% for MvNA; $\sim 1.292$~wt\% for MvNP) represent the median values calculated across all iterations. While any individual model realization strictly yields either excess metal or excess oxygen, but never both, the median values of both quantities can be simultaneously non-zero when the ensemble of Monte Carlo realizations spans both outcomes. For MvNA, both median residuals are small, indicating that the bulk composition lies very close to the boundary of stoichiometric oxygen balance. In contrast, the MvNP results indicate a predominantly oxygen-rich regime.

As a result, when \texttt{ExoInt} partitions iron between the core and mantle based on the available oxygen in the system, the higher oxygen abundance in the MvNP model leads to greater oxidation of iron, stabilizing FeO in the mantle ($\sim$30.52 wt\%).
In comparison, the MvNA model results in a mantle with a moderate FeO content ($\sim$16.27 wt\%).
Substantially high FeO enrichment (beyond $\sim$5-20 wt\%, as seen in the MvNP model) would require core formation to occur under highly oxidizing conditions, in which a larger fraction of iron is retained in the silicate mantle as FeO rather than partitioning into a metallic core \citep{kilburn1997metal, righter2003metal, wade2005core}.
However, such high FeO abundances are generally considered unlikely for typical rocky planets. Instead, core formation is expected to occur under moderately reducing conditions, commonly near the iron-w\"ustite (IW) buffer to approximately IW$-3$, depending on pressure and bulk composition. These conditions are consistent with the mantle FeO abundances observed for Earth ($\sim 8$ wt\%), Mars ($\sim 14$--$18$ wt\%), and some low-pressure differentiated bodies in the inner Solar System ($\sim 18$--$20$ wt\%) \citep{lodders1997oxygen, wade2005core, rubie2011heterogeneous, rubie2015accretion, righter2016metal, doyle2019oxygen}. 
Therefore, the MvNA model, which yields a more balanced distribution of metal and oxygen and maintains physically plausible core--mantle partitioning of iron, is considered to provide a more reliable interior composition for GJ 486b. 
This interpretation is also consistent with the global trend observed across the stellar models, where MvNA generally shows smaller residuals for most elements (see Figure \ref{fig:residual}).

\subsection{Density and Pressure-Temperature Structure} \label{subsec:rho_P_T}

\begin{table*}[ht!]
\centering
\begin{minipage}{0.85\textwidth}
    \centering
    \caption{Bulk mantle compositions in weight percentages (wt\%) of GJ 486b and Earth (Devolatilized and \texttt{HeFESTo}'s model).}
    \label{tab:bulk_composition}
    \setlength{\tabcolsep}{9pt}
    \renewcommand{\arraystretch}{1.4}
    
    \begin{tabular}{|c|c|c|c|c|}
        \hline
        Oxide & Earth$^a$ & Earth & GJ 486b & GJ 486b \\
              & (\texttt{HeFESTo}, wt\%) & (Devolatilization, wt\%) & (MvNP, wt\%) & (MvNA, wt\%) \\
        \hline
        SiO$_2$ & 45.00 & $45.43^{+3.972}_{-4.689}$ & $30.47^{+7.920}_{-4.660}$ & $35.63^{+6.013}_{-6.546}$  \\ 
        Al$_2$O$_3$ & 4.00 & $4.39^{+0.633}_{-0.529}$ & $4.16^{+1.539}_{-0.913}$ & $3.41^{+1.029}_{-0.774}$  \\ 
        FeO & 8.24 & $6.01^{+7.303}_{-4.865}$ & $30.52^{+4.570}_{-12.193}$ & $16.27^{+9.257}_{-9.403}$ \\
        MgO & 38.88 & $39.39^{+5.590}_{-4.770}$ & $30.72^{+12.407}_{-5.161}$ & $39.45^{+11.378}_{-8.231}$ \\ 
        CaO & 3.18 & $3.58^{+0.493}_{-0.428}$ & $2.96^{+1.034}_{-0.645}$ & $3.71^{+1.137}_{-0.832}$  \\ 
        Na$_2$O & 0.13$^*$ & $0.42^{+0.061}_{-0.052}$ & $0.37^{+0.133}_{-0.058}$ & $0.47^{+0.128}_{-0.083}$  \\ 
        Cr$_2$O$_3$$^b$ & 0.57 & $0.78^{+0.120}_{-0.105}$ & $0.80^{+0.282}_{-0.185}$ & $1.07^{+0.329}_{-0.252}$  \\ 
        \hline
    \end{tabular}
    
    \begin{flushleft}
    \small $^a$ Refer Table 2 from \citet{Stixrude-III}. \\
    \small $^b$ Cr has been incorporated by extrapolating its devolatilization coefficient from those of Fe and Si based on their $T_{\mathrm{c},50}$ values, following the methodology of \citet{wang2022detailed}. \\ 
    \small $^*$ Na$_2$O is exceptionally low here. Refer to Table 3 from \citet{workman2005major}.
    \end{flushleft}
\end{minipage}
\end{table*}

The Earth and GJ 486b show modest differences in their interior structures. While Earth has a mantle thickness of $\sim2900$ km (0.462 $R_{\oplus}$), GJ 486b has a thicker mantle of $\sim3966$ km (0.463 $R_{p}$). 
We note that the core radius fraction and core-mantle boundary (CMB) depth for GJ~486b adopted here are derived following \citet{sahu2025unveiling}, who assume an Earth-like baseline core mass fraction and core light-element composition (20\%; 12\% silicon, 7\% oxygen, and 1\% sulfur). In reality, variations in the bulk planetary Fe/Si ratio, mantle Fe/FeO partitioning, or core light-element fraction would shift the CMB depth and pressure. For example, a higher bulk Fe/Si ratio would reduce the mantle thickness and alter the CMB pressure, whereas variations in Fe/FeO partitioning would further modify the mantle structure and pressure profile. While this assumption may not be generally applicable, developing a fully self-consistent model that simultaneously accounts for variations in bulk composition, core-mantle partitioning, and core light-element composition is beyond the scope of this study.

Owing to its larger size, the mantle density of GJ 486b reaches up to $\sim1.5\times$ that of Earth (see Figure 4 in \citet{sahu2025unveiling}).
The pressure--depth profile of GJ 486b also exceeds that of Earth in our calculations, with the core--mantle boundary located at approximately 390 GPa for GJ 486b, compared to about 138 GPa for Earth.
As a result, the adiabatic $P$--$T$ profiles show a noticeable divergence between Earth and GJ 486b.

\begin{figure}[ht!]
    \includegraphics[width=\linewidth,page=1]{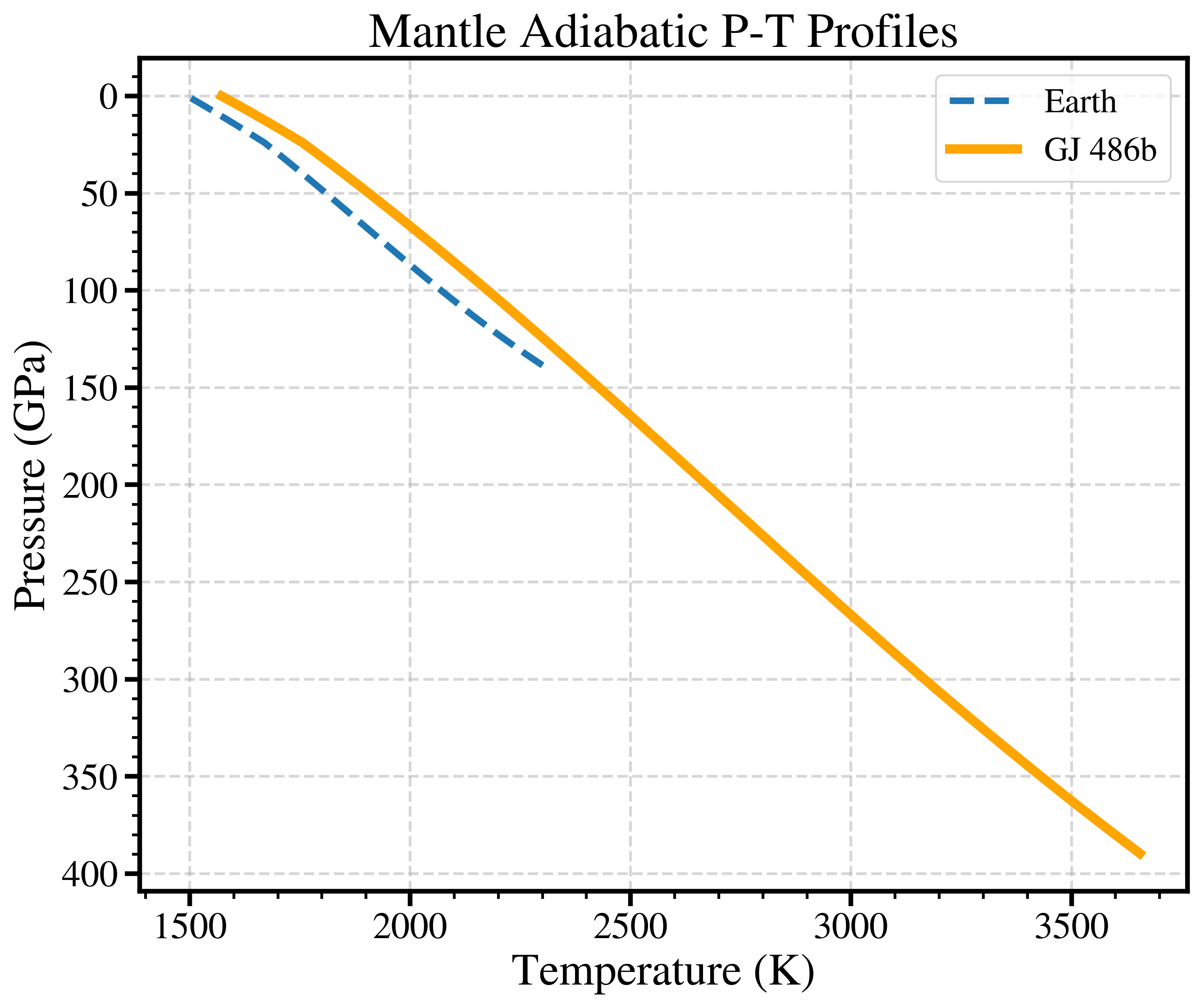}
   \caption{Comparison of mantle adiabatic $P$--$T$ profiles for Earth and the exoplanet GJ 486b.}  \label{fig:PT_planets}
\end{figure}

We use the $P$--$T$ profile of the planet after the thermal evolution is complete and the planet enters a cool post-magma-ocean state. 
The $P$--$T$ profile (shown in Figure \ref{fig:PT_planets}) represents the mantle adiabat at a mantle potential temperature of 1572 K.
It indicates that the mantle of GJ 486b experiences extreme pressure regimes even at relatively shallow depths, which may stabilize high-pressure polymorphs such as bridgmanite, ferropericlase, post-perovskite, and majorite over a broader depth range than on Earth \citep{murakami2004post, Boukare2015}. These phases extend to greater depths, with post-perovskite stabilising over a substantially broader pressure range than in the relatively thin D$''$ layer, where post-perovskite occurs on Earth \citep{murakami2004post}. The higher interior pressures of GJ~486b thus extend the stability fields of lower-mantle phases well beyond the pressure conditions accessible in Earth's interior \citep{Stixrude2012}.


\subsection{Bulk Mantle Composition} \label{subsec:bulk_mantle}
 
Table \ref{tab:bulk_composition} summarizes the bulk mantle composition of Earth and GJ 486b from our models and references values mentioned in \citet{Stixrude-III}.
Our results indicate that GJ 486b is substantially Fe-rich. The MvNP and MvNA models show FeO contents of $\sim30.52$ wt\% and $\sim16.27$ wt\%, respectively, compared to $\sim8.24$ wt\% and $\sim6.01$ wt\% for Earth, as reported by
\citet{workman2005major} and predicted by the devolatilised Earth model


In contrast, SiO$_2$ is significantly depleted in GJ 486b, with values of $\sim30.47$ wt\% and $\sim35.63$ wt\% for the MvNP and MvNA models, respectively, whereas Earth shows higher values of $\sim45$ wt\% in both models.
The MgO abundance in the MvNA model ($\sim39.45$ wt\%) remains close to terrestrial values but decreases significantly in the MvNP model ($\sim30.72$ wt\%).
Also CaO shows some variation, with values of $\sim2.96$ wt\% and $\sim3.71$ wt\% for the MvNP and MvNA models of GJ 486b, respectively.
For Al$_2$O$_3$, MvNA model produces a significantly lower amount than the terrestrial values ($\sim3.41$ wt\%), but MvNP produces a value which falls in between the two terrestrial values in the table ($\sim4.16$ wt\%).
Other minor components like Na$_2$O and Cr$_2$O$_3$ also vary slightly.
For both models, the Na$_2$O value remains closer to the devolatilized mantle composition ($\sim0.37$ wt\% in MvNP and $\sim0.47$ wt\% in MvNA).
However, Cr$_2$O$_3$ value increases slightly in both models of GJ 486b ($\sim0.8$ wt\% in MvNP and $\sim1.07$ wt\% in MvNA). 
These differences are also evident in the top panel of Figure \ref{fig:Predicted_elemental_abundances} and in the ternary diagrams shown in Figure \ref{fig:ternary_plots}.


The reduced SiO$_2$ and enhanced FeO contents of GJ~486b arise primarily from the super-solar Mg/Si ratio of its host star ($[\text{Mg/Si}] = +0.10$~dex). Because MgO and SiO$_2$ require one and two oxygen atoms per Mg and Si atom, respectively, the relatively lower Si abundance leaves more oxygen available that would otherwise be consumed in forming SiO$_2$. This excess oxygen can instead oxidise Fe to form FeO, a mechanism previously identified by \citep{unterborn2017effects}. This effect is further supported by the devolatilisation corrections and oxygen mass-balance partitioning framework adopted in \texttt{ExoInt} \citep{wang2019enhanced}.


In traditional petrological models, the oxidation state is specified by selecting a buffer that fixes the ratio of FeO to metallic Fe at a given pressure and temperature. In contrast, \texttt{ExoInt} estimates iron partitioning using the oxygen mass-balance framework described in Section \ref{subsec:inferring_bulk_composition}. Within this framework, the FeO abundance and core mass fraction are determined by the available oxygen budget inherited from the devolatilized stellar composition rather than by prescribing a fixed oxygen fugacity buffer \citep{wang2022detailed}. The higher oxygen availability in the MvNP composition, therefore, results in greater abundance of FeO within the mantle, whereas the lower oxygen budget in the MvNA composition leads to increased retention of metallic iron.

This distinct bulk chemistry directly impacts mineral stability.
Since phase assemblages are computed through Gibbs free energy minimization at fixed bulk compositions, increasing bulk FeO modifies the thermodynamic equilibrium conditions, leading to changes in the stable phase assemblage and in the distribution of Fe among silicate and oxide solid solutions.
Increased FeO stabilizes ferropericlase and Fe-bearing bridgmanite while reducing the relative stability of silica-rich phases such as stishovite. Consequently, the mineralogical differences between Earth and GJ 486b follow systematically (see Figures \ref{fig:Earth_HeFESTo_devol} and \ref{fig:mvna_mvnp_final_gj}), arising primarily from the Fe/Mg/Si shifts imposed at the bulk composition stage.

\subsection{Benchmarking Earth against \texttt{HeFESTo}} 
\label{subsec:validation_HeFESTo}

\begin{figure*}[ht!] 
    \centering
    \includegraphics[width=\linewidth]{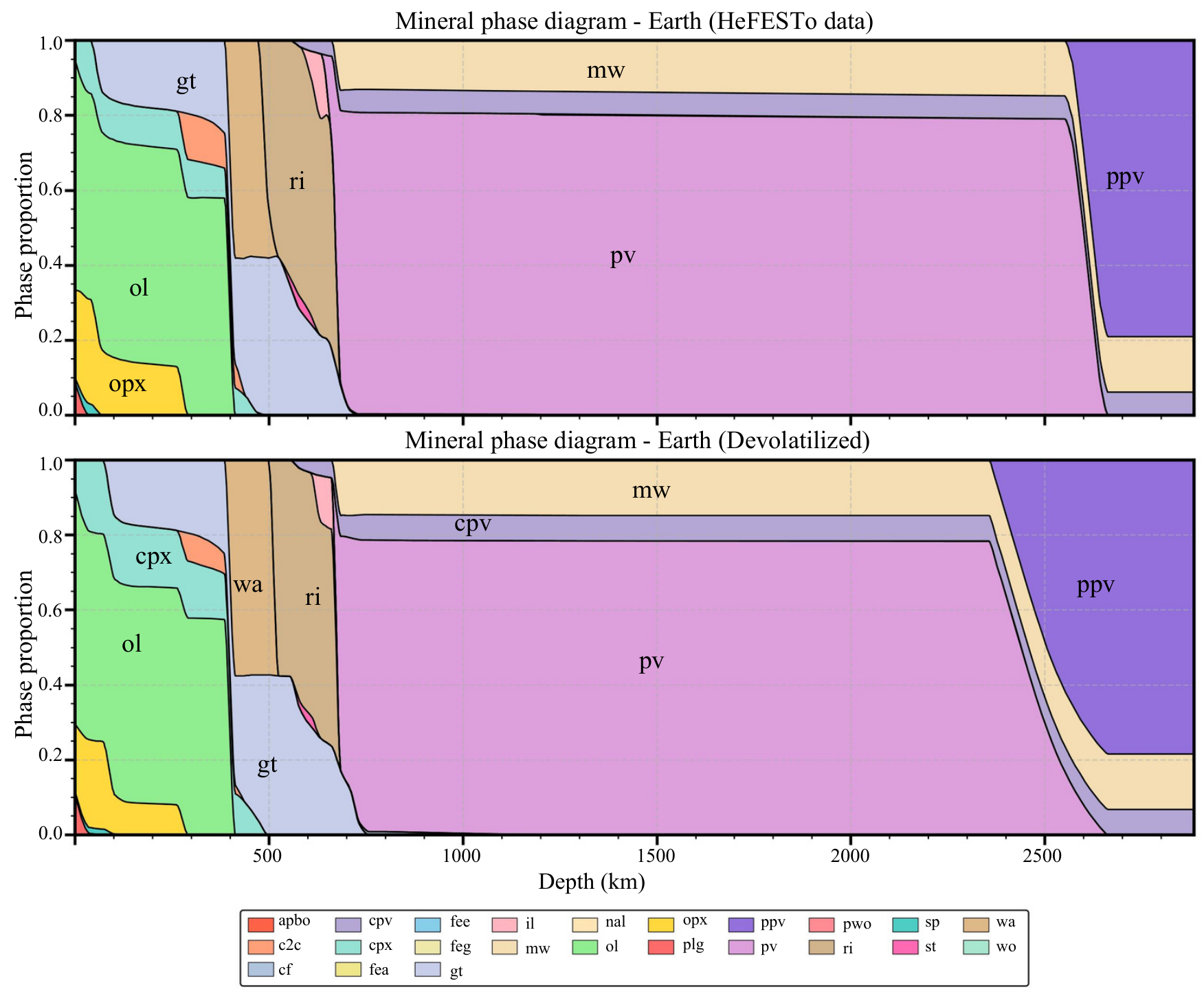}
   \caption{Comparison of Earth's mantle phase assemblages as a function of depth (x-axis) for two different bulk composition prescriptions. Top panel: phase proportions (in wt\%) obtained using bulk composition and phase-equilibrium calculations from \texttt{HeFESTo}. Bottom panel: phase proportions (in wt\%) obtained using a bulk composition derived by applying the devolatilization model to solar abundances.}
    \label{fig:Earth_HeFESTo_devol}
\end{figure*}

 We first perform a consistency check on the results obtained from the ESH framework by comparing the predicted Earth's mantle mineralogy with a reference phase diagram generated independently using \texttt{HeFESTo} \citep{Stixrude-II}.
This reference model serves as a benchmark for Earth's mantle and has been widely adopted and tested in previous studies \citep{Stixrude-II, Stixrude-III}.
The devolatilization model of \texttt{ExoInt} is calibrated using the Earth-Sun system and Solar System data, reproducing 
a particular model of Earth's composition as a proof of concept \citep{wang2019volatility, wang2019enhanced}. 
The default oxide composition in \texttt{HeFESTo} (the depleted upper mantle composition from \citet{workman2005major}, given in Table \ref{tab:bulk_composition}) differs slightly from that adopted in the \texttt{ExoInt} calibration. 
This discrepancy may lead to small differences in the predicted baseline phase assemblages. 

Figure \ref{fig:Earth_HeFESTo_devol} compares mantle phase diagrams for the benchmark and our model for Earth.
The upper panel of Figure \ref{fig:Earth_HeFESTo_devol} shows the reference phase diagram for Earth, generated using the mantle composition provided in \texttt{HeFESTo} by default and the corresponding phase equilibrium calculations performed using \texttt{HeFESTo} \citep{Stixrude-II}. Hereafter, this is referred to as the \texttt{HeFESTo}-only framework.
The lower panel of Figure \ref{fig:Earth_HeFESTo_devol} shows the phase diagram for Earth in which the bulk composition is first calculated using the MvNP stellar model and then the ESH framework is applied.
As both phase diagrams are constructed using the same thermodynamic framework, any differences in mineral  stability domains and modal proportions can be directly attributed to variations in bulk major oxide composition rather than to differences in phase equilibrium methodology
Hence, this comparison provides a robust test of whether the compositional modifications introduced by the ESH framework preserve the first-order mineralogical structure of Earth's mantle.

The phase diagrams reveal that the broad mineralogical architecture of the mantle remains consistent between the \texttt{HeFESTo}-only and ESH frameworks, although the relative  stability domains and proportions of the major phases show some systematic differences. In the \texttt{HeFESTo}-only results, the upper mantle is dominated by olivine, orthopyroxene, clinopyroxene, and garnet, with broadly comparable  stability regimes  for orthopyroxene and clinopyroxene.
 In this model, the phase proportion of clinopyroxene at all depths is slightly smaller while orthopyroxene is slightly enriched. 
As depth increases, olivine undergoes its first major polymorphic transition to wadsleyite near 410 km, followed by the transition from wadsleyite to ringwoodite around 500 km, marking the progression through the transition-zone assemblages.
Pyroxenes also undergo breakdown across this interval ($\sim410$--660 km), producing majoritic garnet, ilmenite, and other associated high-pressure silicate phases.

In the ESH framework results, the upper mantle is characterized by the same dominant mineral assemblage of olivine, orthopyroxene, clinopyroxene, and garnet.
However, subtle differences in modal proportions are observed. Orthopyroxene stability domain decreases by a small amount, while clinopyroxene occupies slightly higher phase proportion than the benchmark model.
The olivine and garnet field narrows moderately.
Despite these differences, the sequence of olivine polymorphic transitions is preserved, with the olivine--wadsleyite transition occurring near $\sim410$ km and the wadsleyite--ringwoodite transition near $\sim500$ km.
Pyroxenes similarly undergo breakdown across the transition zone, producing the same phases over comparable depth ranges.

Within Earth's transition zone, the benchmark model shows well-defined wadsleyite and ringwoodite stability ranges, accompanied by the development of ilmenite and other high-pressure silicate phases.
The same can also be observed in the ESH framework, as shown in Fig. \ref{fig:Earth_HeFESTo_devol}. 
In the \texttt{HeFESTo}-only results, ringwoodite breaks down to form bridgmanite (Mg-perovskite), calcium-perovskite and ferropericlase, establishing the lower mantle assemblage.
This region remains dominated by bridgmanite with subordinate ferropericlase and calcium-perovskite  throughout most of the lower mantle. 
At greater depths, bridgmanite undergoes its final major transformation into post-perovskite, typically around 2600--2700 km depth, marking the deepest mineralogical transition.
The ESH framework reproduces the same lower mantle phase sequence, including the formation of bridgmanite and ferropericlase and the appearance of post-perovskite in the lowermost mantle. However, the devolatilized composition results in breakdown of bridgemanite at a relatively shallower depth to form post-perovskite at around 2300 km depth.
As a result, the bridgmanite field gets reduced while post-perovskite occupies a slightly larger region of the figure in the lower mantle compared to the \texttt{HeFESTo}-only framework.
The stability regions of magnesiow\"ustite and calcium perovskite also increase slightly. 

Overall, the slight mineralogical differences between the results obtained from the \texttt{HeFESTo}-only and ESH frameworks reflect small increases in bulk Mg/Fe and lower incompatible element content, which modestly adjust the stability of olivine, pyroxenes, bridgmanite, ferropericlase, and related phases.
These trends are consistent with experimental high-pressure studies \citep{fei2004experimentally, murakami2004post, hirose2006postperovskite}.
Importantly, the major phase sequence, transition pressures, and overall mantle structure remain effectively unchanged, indicating that both models produce broadly similar mineralogical profiles despite these subtle compositional shifts.

\subsection{Mineralogy of GJ 486b and Comparison with Earth} \label{subsec:GJ_Earth}

\begin{figure*}[ht!]
    \centering
    \includegraphics[width=\linewidth]{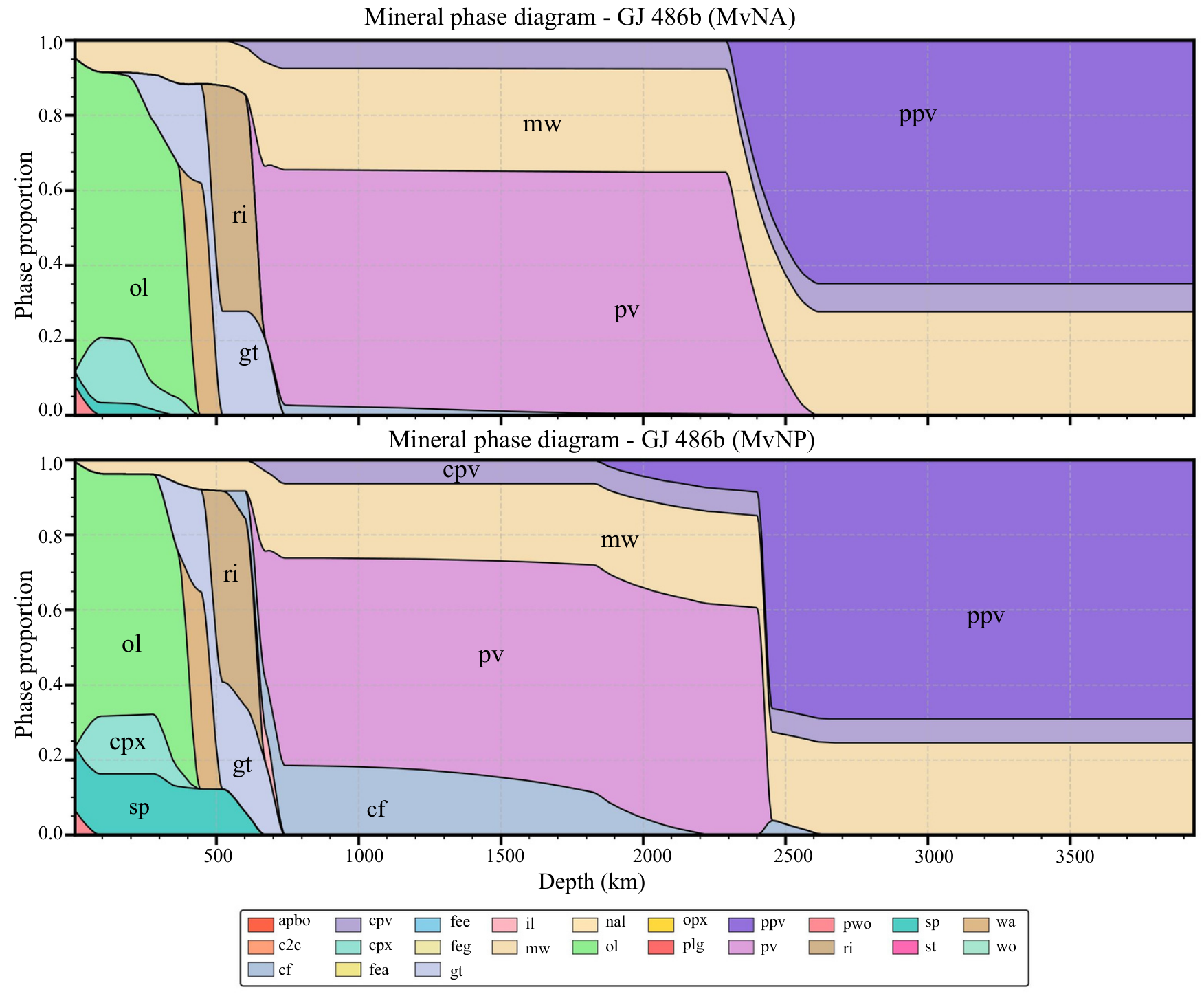}
   \caption{Phase assemblages of GJ 486b as a function of depth derived from two stellar abundance inference approaches. The MvNP model (top), conditioned only on stellar parameters, and the MvNA model (bottom), which incorporates observed abundance correlations, both predict a strongly olivine-rich upper mantle and a lower mantle dominated by bridgmanite, with post-perovskite stable in the lowermost mantle. Variations in phase proportions (wt\%) and transition depths between the two cases highlight how uncertainties in inferred stellar abundances propagate into the predicted interior mineralogy.}
    \label{fig:mvna_mvnp_final_gj}
\end{figure*}

Figure \ref{fig:mvna_mvnp_final_gj} shows the mineral phase diagram for GJ 486b, revealing a mantle mineralogy extending to substantially higher pressures than that of Earth (see Figure 4 of \citet{sahu2025unveiling}). 
Although the overall sequence of silicate phase transitions broadly resembles that of Earth, notable differences are observed in the stability domains and modal abundances of major minerals.
In GJ 486b, mineralogical transitions occur over shorter depth intervals, high-pressure phases stabilize at relatively shallower depths, and several phases show either expanded or reduced stability domains relative to Earth.
These systematic shifts reflect the steeper pressure gradient and stronger gravity characteristic of super-Earth interiors \citep{valencia2006internal, seager2007mass}.

While the steeper pressure gradient of GJ 486b promotes earlier stabilization of high-pressure polymorphs, the elevated bulk FeO and reduced SiO$_2$ content shift equilibrium toward Fe-rich phases, expanding the stability domain of magnesiow\"ustite and reducing the relative abundance of bridgmanite.
Thus, the observed mineralogical shifts arise from a coupled effect of pressure scaling and bulk compositional modification.

\subsubsection{Upper Mantle Assemblage} \label{subsubsec:upper_mantle}

Our results show that, on GJ 486b, the upper mantle is dominated by olivine, accompanied by spinel, clinopyroxene, garnet, magnesiow\"ustite, and plagioclase (Figure \ref{fig:mvna_mvnp_final_gj}).
This general mineral assemblage is similar to that of Earth, and the sequence of low-pressure silicate phases follows the terrestrial upper mantle mineralogy predicted by thermodynamic models.

However, notable differences arise in the relative stability regimes and modal abundances. Our results indicate a smaller clinopyroxene stability zone in GJ 486b compared to Earth's mantle, along with a broadened spinel and olivine stability zone for both the MvNP and MvNA model. 
These changes reflect the onset of high-pressure effects at relatively shallow depths in the mantle of GJ 486b.

\subsubsection{Transition Zone Assemblage} \label{subsubsec:transition_zone}

Within the transition zone of GJ 486b, olivine undergoes rapid polymorphic transformations to wadsleyite and subsequently to ringwoodite over a narrow depth interval. This zone is primarily composed of assemblages bearing garnet, ilmenite, wadsleyite, and ringwoodite, along with small amounts of calcium-ferrite.

The overall sequence of olivine polymorphic transitions remains consistent with that of Earth, indicating similar phase relations.
However, the garnet stability zone is significantly reduced, and both the stability domain and the modal abundance of ringwoodite are decreased.
Although the wadsleyite zone spans a similar depth range to that on Earth, its modal abundance is lower in GJ 486b.
We also observe a noticeable amount of ilmenite in the MvNP model, which is negligible in the MvNP model. However, both of these show reduced abundances relative to Earth's mantle.
In addition, the calcium-ferrite structured phase becomes noticeable in case of  GJ 486b, especially in the MvNP model, which was negligible for the case of Earth. 
These changes indicate a transition-zone mineralogy characterized by different proportions of intermediate-pressure phases.

\subsubsection{Lower Mantle Assemblage} \label{subsubsec:lower_mantle}

Below the transition zone, ringwoodite breaks down to form a lower-mantle assemblage dominated by bridgmanite, magnesiow\"ustite, calcium perovskite, and calcium-ferrite.
This assemblage is broadly similar to Earth's lower mantle in terms of mineral sequence and phase identity.

However, several systematic differences are evident in this region.
The stability domain of bridgmanite is significantly shrunk.
Since bridgmanite has the stoichiometry (Mg,Fe)SiO$_3$, its abundance is limited by the available SiO$_2$ in the bulk mantle.
The lower SiO$_2$ content of GJ 486b therefore restricts bridgmanite formation and favors the formation of magnesiow\"ustite.
As a result, magnesiow\"ustite shows occupies a larger region in the lower mantle.
The calcium-ferrite structured phase also shows a broad stability zone in GJ 486b (specially for MvNP model). 
Calcium perovskite remains stable over a depth range comparable to that on Earth, but its modal abundance is increased. These shifts indicate redistribution of modal proportions (as mentioned in Table \ref{tab:bulk_composition}) among the lower-mantle phases under higher pressure conditions.

At greater depths, bridgmanite transforms into post-perovskite, producing a mantle assemblage dominated by post-perovskite together with magnesiow\"ustite and calcium perovskite.
While this phase transition is also present on Earth, its expression differs in GJ 486b. 
In case of MvNA model, the transition of bridgemanite to post-perovskite occurs at a depth comparable to that of Earth ($\sim$ 2300 km), whereas in the MvNP model, it occurs at a much shallower depth ($\sim$ 1800 km) reflecting the effect of high pressure and composition changes at shallower depths.
Compared to Earth, the modal abundance of post-perovskite in GJ 486 b is lower at any given depth. This is a direct consequence of the silica-limited bridgmanite stability field and the corresponding expansion of the magnesiow\"ustite field. 
However, because GJ 486b possesses a substantially larger mantle than Earth, the total amount of post-perovskite in its lower mantle is greater despite its lower modal abundance.

Therefore, the results show that, while GJ 486b and Earth share a broadly similar sequence of silicate phase transitions, the mantle of GJ 486b shows systematic differences in depth distribution and modal abundances.
Relative to Earth, GJ 486b displays slightly smaller stability domain and abundances of pyroxenes, garnet, ringwoodite, and bridgmanite, alongside expanded stability domains of spinel, magnesiow\"ustite, calcium-ferrite, and calcium perovskite.
These mineralogical differences reflect the steeper pressure-temperature gradient of GJ 486b, which promotes earlier stabilization of high-pressure phases, as well as bulk compositional differences that favor Fe-bearing phases \citep{valencia2006internal, Unterborn_2018, unterborn2019pressure}.

\subsection{P-T Sensitivity of Mantle Mineralogy}
\label{subsec:sensitivity_P_T}


\begin{figure*}[ht!] 
    \centering    \includegraphics[width=\linewidth]{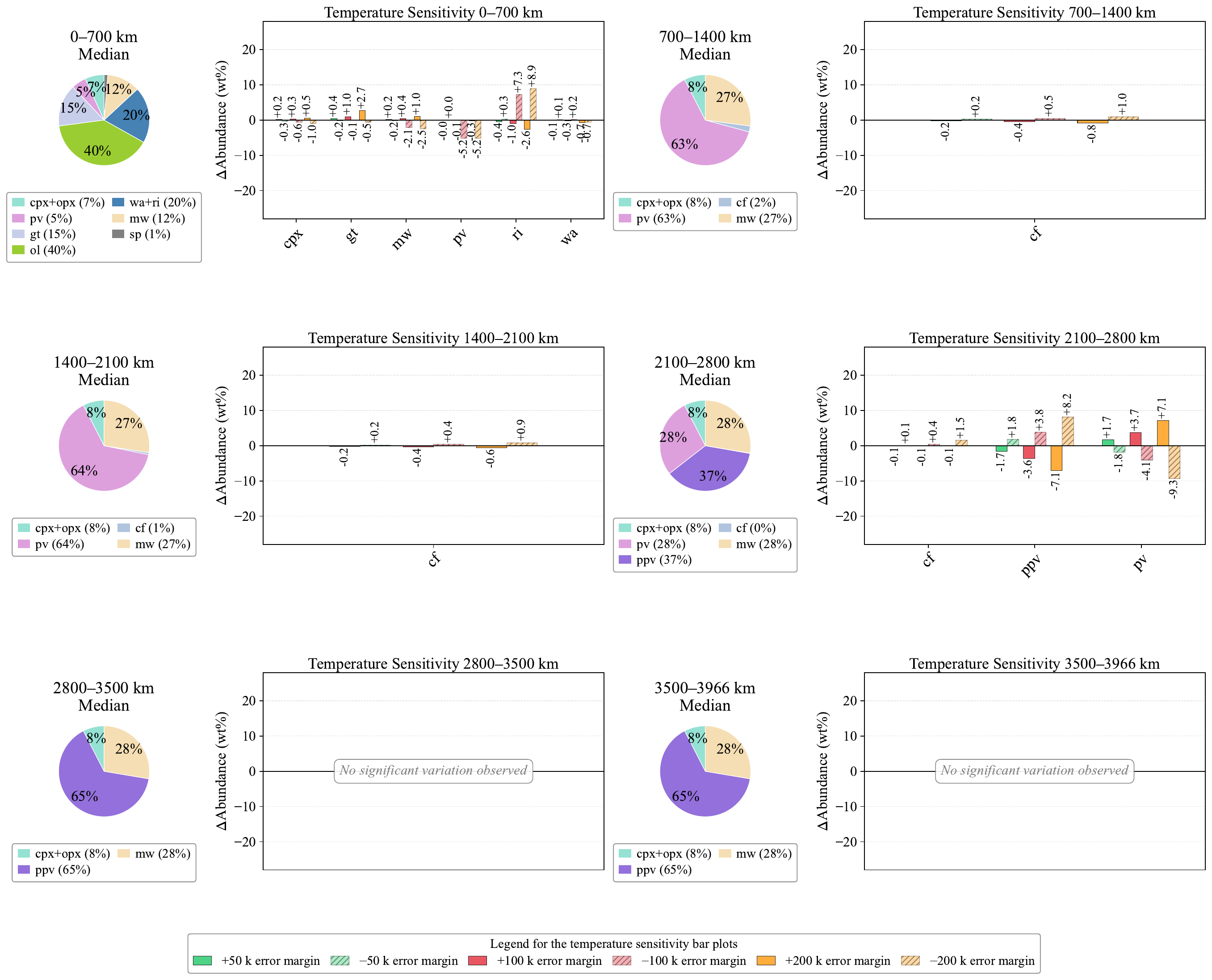}
    \caption{Temperature sensitivity of the predicted mantle mineral assemblage from the MvNA model. Each row corresponds to a 700 km depth bin spanning the full depth range of 0--3966 km. Within each row, the left panel shows a pie chart of the median mineral assemblage (in wt\%) for that depth bin, with mineral groups color-coded and labeled. The right panel shows the change in phase abundance ($\Delta$ abundance, wt\%) relative to the median for temperature perturbations of $\pm 50$ K, $\pm 100$ K, and $\pm 200$ K applied uniformly along the reference pressure-temperature path. Solid bars denote positive perturbations ($+\Delta T$), and hatched bars denote negative perturbations ($-\Delta T$); bar pairs are color coded by perturbation magnitude as indicated in the legend. 
    Only phases whose relative change exceeds 10\% of the median abundance are shown.}
    \label{fig:MvNA_temp_sensitivity}
\end{figure*}

To test the robustness of the obtained mineral phase proportions for GJ 486b, we perturbed the $P$--$T$ profile by varying the mantle potential temperature by $\pm50$ K, $\pm100$ K, and $\pm200$ K, while keeping the bulk composition fixed.
These calculations were performed separately for both the MvNP and MvNA models.

\begin{figure*}[ht!] 
    \centering
    \includegraphics[width=\linewidth]{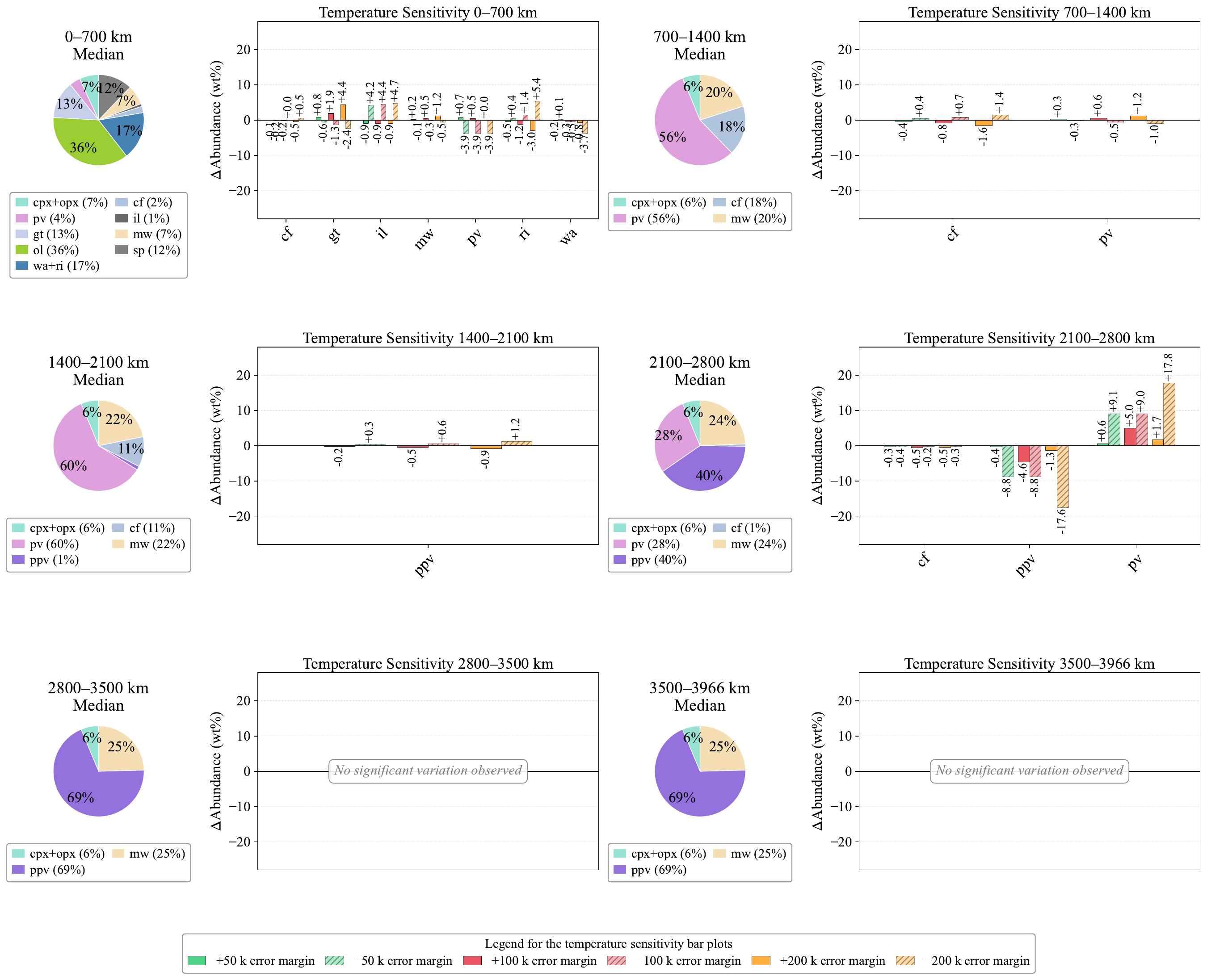}
    \caption{Temperature sensitivity of the predicted mantle mineral assemblage from the MvNP model. Each row corresponds to a 700 km depth bin spanning the full depth range of 0--3966 km. Within each row, the left panel shows a pie chart of the median mineral assemblage (mass fraction, wt\%) for that depth bin, with mineral groups color-coded and labeled. The right panel shows the change in phase abundance ($\Delta$ abundance, wt\%) relative to the median for temperature perturbations of $\pm 50$ K, $\pm 100$ K, and $\pm 200$ K applied uniformly along the reference pressure-temperature path. Solid bars denote positive perturbations ($+\Delta T$), and hatched bars denote negative perturbations ($-\Delta T$); bar pairs are color coded by perturbation magnitude as indicated in the legend. 
    Only phases whose relative change exceeds 10\% of the median abundance are shown.}
    \label{fig:MvNP_temp_sensitivity}
\end{figure*}

\ref{fig:MvNA_temp_sensitivity} and \ref{fig:MvNP_temp_sensitivity} show the resulting modal abundances and their deviations from the baseline model (bar plots) for mineral phases that show significant variations in the MvNA and MvNP abundance models, respectively.
For the cases where deviations are below 0.1\%, we consider the variations to be negligible.

For the MvNP model, the temperature variations have little effect on the mineralogy at the upper portion of the lower mantle (700-2100 km), whereas noticeable shifts can be observed at the shallower (0-700 km) and deeper (2100-2800 km) depths. Again, at much deeper parts, no variations can be seen for the temperature perturbation.
For temperature offsets of up to $\pm50$ K, the phase assemblage remains almost unchanged for most pressure levels.
Though minor redistribution occurs among coexisting silicates, the dominant phases at each depth remain stable, and the modal differences stay well below the level required to alter the overall mineralogical structure. 

At $\pm100$ and $\pm200$ K, variations become more apparent.
Moderate adjustments in the proportions of high-pressure silicates emerge, reflecting the temperature dependence of their stability zones, yet the underlying assemblage remains unchanged.
The variations are generally below 10\% for most phases except bridgemanite and post-perovskite (2100-2800 km depth interval).
The iron-bearing phases, such as garnet, ilmenite, ringwoodite, bridgemanite, and post-perovskite are showing the largest variations.
In the upper mantle region, where multiple phases coexist, the perturbations produce only small shifts in relative abundances rather than introducing new stable phases.

The strongest sensitivity occurs near the perovskite to post-perovskite transition (2500--3000 km).
Here, noticeable redistribution between perovskite and post-perovskite is produced due to changes in $P$--$T$ conditions for all three temperature perturbation conditions.
However, this effect is confined to a narrow depth range near $\sim2600$ km, associated with the shift of the pv--ppv transition.
At greater depths, the mantle stabilizes into a dominant phase field, and modal abundances again become largely insensitive to temperature perturbations.


In the case of the MvNA model, temperature variations of $\pm50$ K have little effect on the mineralogy throughout the mantle, and noticeable shifts begin to appear only when the perturbation reaches $\pm100$ and $\pm200$ K. 
Though a similar trend can also be seen for this model, the variations are always below 10\% throughout the mantle.



The purpose of the $P$--$T$ sensitivity analysis text is to assess whether the use of Earth-calibrated density relations to construct the interior profile of GJ 486b introduces significant mineralogical bias.
Although the bulk mantle composition derived from \texttt{ExoInt} differs from that of Earth, the structural calculation within \texttt{SERPINT} assumes an Earth-like density framework to generate the pressure profile.

Our results show that, for temperature variations of up to $\pm200$ K, the dominant mantle phase assemblage remains stable across most of the interior (as shown in Figures \ref{fig:MvNA_temp_sensitivity} and \ref{fig:MvNP_temp_sensitivity}).
The greatest sensitivity is confined to narrow depth regions near major phase boundaries, particularly the perovskite to post-perovskite transition.
Outside these boundary regions, modal abundances vary only modestly.
This indicates that reasonable uncertainties in the adopted $P$--$T$ profile do not significantly alter the first-order mineralogical structure of GJ 486b.
The dominant control on mantle mineralogy arises from bulk composition, while temperature primarily modulates transition depths rather than phase identity.

\subsection{Sensitivity analysis of mineralogy against bulk compositional and stellar abundance models}

The stellar abundances derived from \textit{Gaia} DR3 carry observational uncertainties that propagate through the \texttt{ExoInt} and \texttt{HeFESTo} workflow, leading to uncertainties in the predicted mantle mineralogy. To assess the robustness of our results, we perform a sensitivity analysis using the upper and lower uncertainty bounds of the bulk compositions derived from the MvNP and MvNA stellar abundance models.

The median bulk compositions for the MvNP and MvNA models are listed in columns 4 and 5 of Table \ref{tab:bulk_composition}, respectively. The corresponding upper and lower uncertainty bounds are applied to these median compositions, and the resulting equilibrium phase assemblages are calculated using the ESH framework, keeping the $P$--$T$ profile fixed. 
In Figures \ref{fig:MvNP_composition_sensitivity} and \ref{fig:MvNA_composition_sensitivity}, we show the sensitivity plot for mineral phases that show variations greater than 10\% relative to the median values.

\subsubsection{Sensitivity to composition in the MvNP model}
\label{subsec:sensitivity_mvnp}

\begin{figure*}[ht!] 
    \centering
    \includegraphics[width=\linewidth]{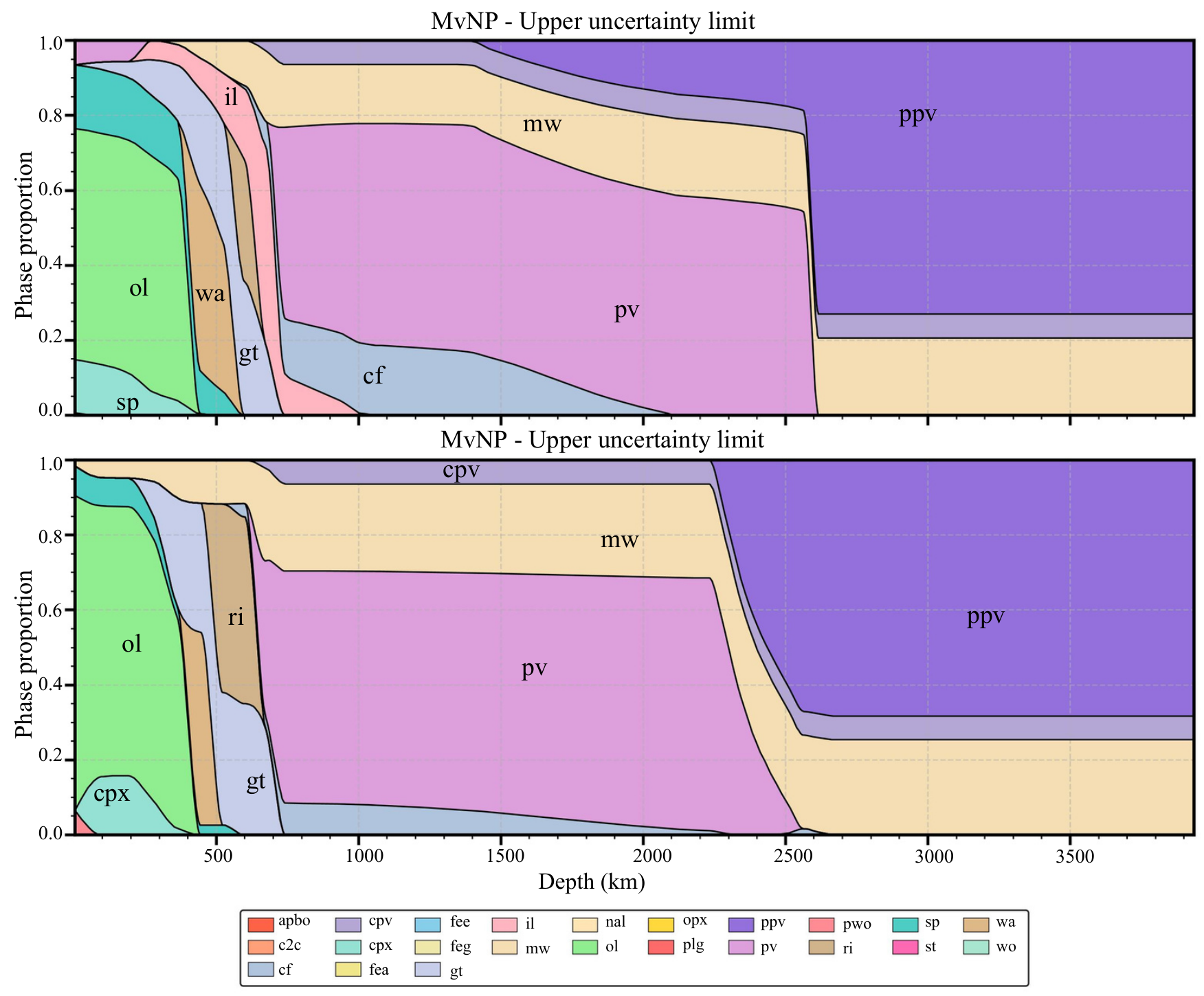}
  \caption{Phase proportion (in wt\%) profiles as a function of depth for the MvNP compositional model under stellar abundance perturbations. The top panel shows the positive oxide perturbation and the bottom panel shows the negative oxide perturbation relative to the median model. Phase proportions are normalized at each depth.}
    \label{fig:err_MvNP_phase}
\end{figure*}

\begin{figure*}[ht!] 
    \centering
    \includegraphics[width=\linewidth]{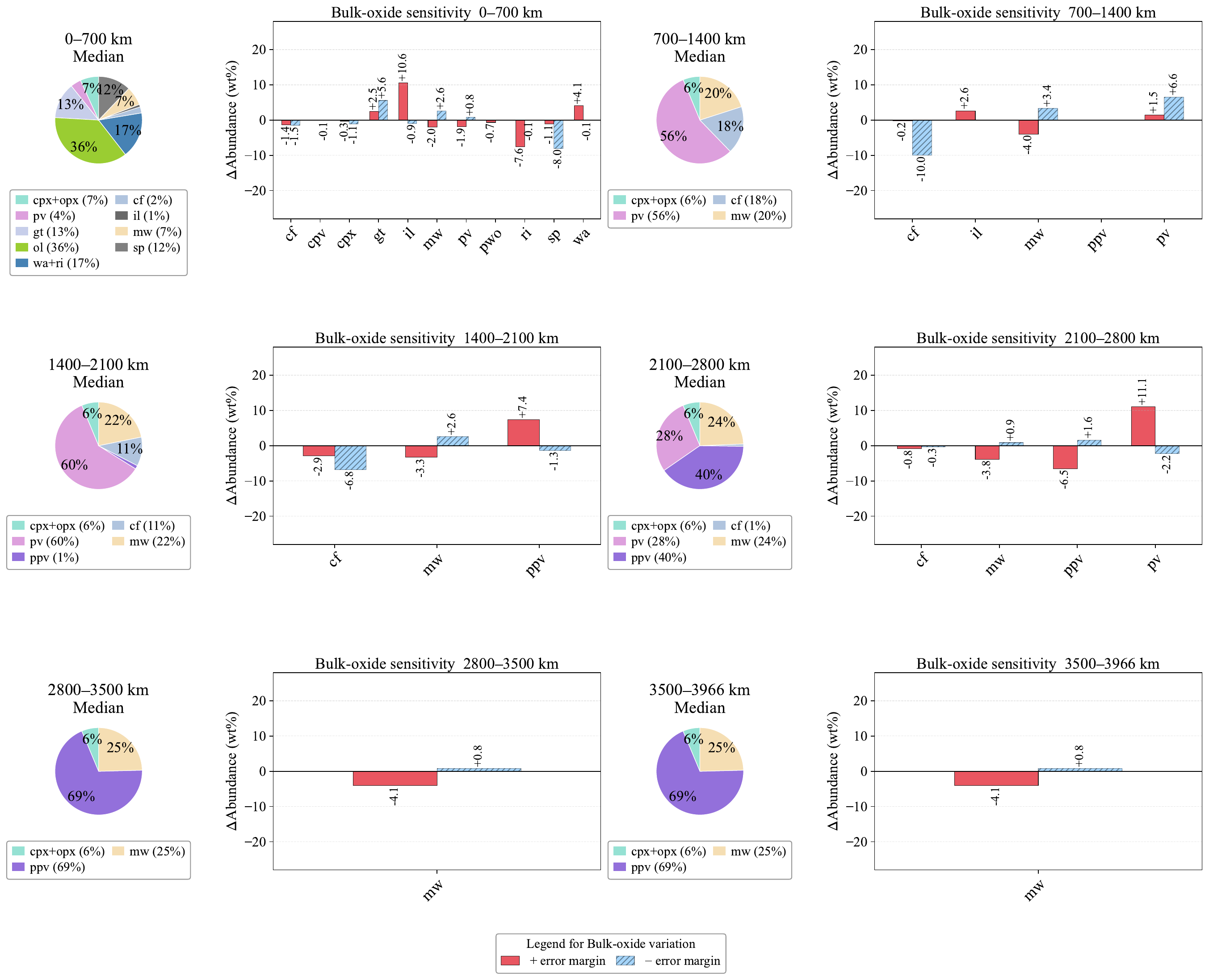}
  \caption{Sensitivity of mantle mineralogy to variations in composition corresponding to the upper and lower uncertainty bounds derived from the MvNP model. The data are presented in Table \ref{tab:bulk_composition}. 
Under compositional perturbations, the mantle mineralogy shows relatively small variations. These variations are primarily confined to the lower mantle and transition zone, particularly in magnesiow\"ustite, bridgmanite (perovskite), and post-perovskite.}
    \label{fig:MvNP_composition_sensitivity}
\end{figure*}

Figure \ref{fig:err_MvNP_phase}(a) and (b) show the mineral phase diagrams corresponding to the compositions at the upper and lower uncertainty limits, respectively, for the MvNP model, compared to the median case shown in Figure \ref{fig:mvna_mvnp_final_gj}(a).
To quantify these variations, we compute the percentage change in mineral phases that are significantly affected by the uncertainty ranges.

As discussed in Sections \ref{subsec:stellar_abundance_models}, \ref{subsec:inferring_bulk_composition}, and \ref{subsec:bulk_mantle}, the primary factor driving the large compositional variations is the choice of the stellar abundance model.
The MvNP model is conditioned on stellar parameters and reflects variations across different stellar properties in the catalog.
The largest compositional variations occur for Mg, Fe, and Si. As a result, the abundances of associated Fe--Mg silicate minerals vary significantly across the uncertainty ranges (see Figure \ref{fig:err_MvNP_phase}).

In the upper mantle, the variations remain below 11\%, with only changes observed in garnet, ilmenite, spinel, ringwoodite and related minerals.
Sensitivity increases progressively with depth.
As depth increases into the lower mantle, larger variations are observed for bridgmanite (perovskite), post-perovskite, magnesiow\"ustite, and calcium-ferrite (as shown in Figure \ref{fig:err_MvNP_phase}).
Because the lower mantle is dominated by Mg-, Fe-, and Si-bearing phases, variations in these elemental abundances produce stronger mineralogical variations than in the upper mantle.

\subsubsection{Sensitivity to composition in the MvNA model}
\label{subsec:sensitivity_mvna}

\begin{figure*}[ht!] 
    \centering
    \includegraphics[width=\linewidth]{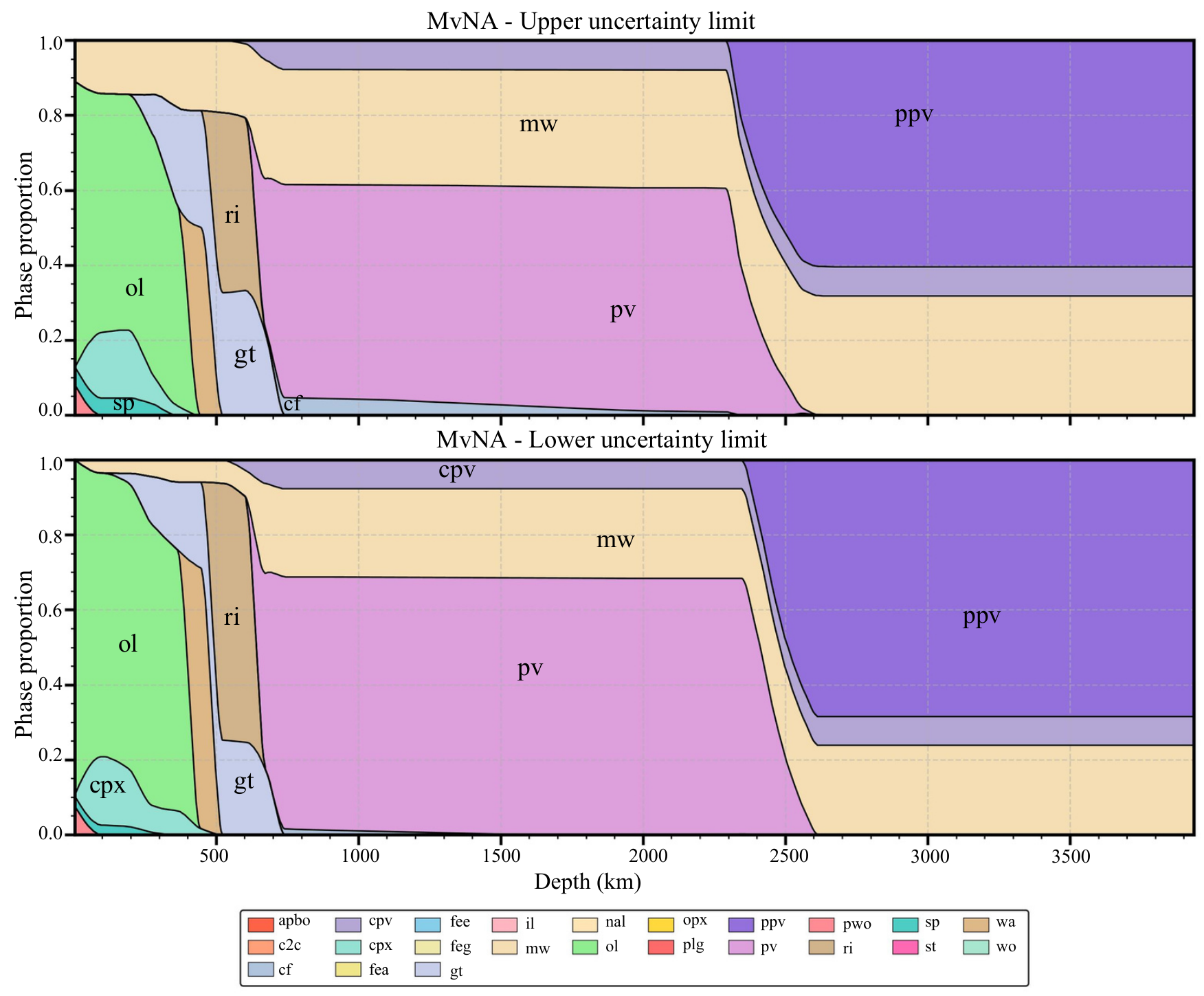}
    \caption{Phase proportion (in wt\%) profiles as a function of depth for the MvNA compositional model under stellar abundance perturbations. The top panel shows the positive oxide perturbation, and the bottom panel shows the negative oxide perturbation relative to the median model. Phase proportions are normalized at each depth.}
    \label{fig:err_MvNA_phase}
\end{figure*}

\begin{figure*}[ht!] 
    \centering
    \includegraphics[width=\linewidth]{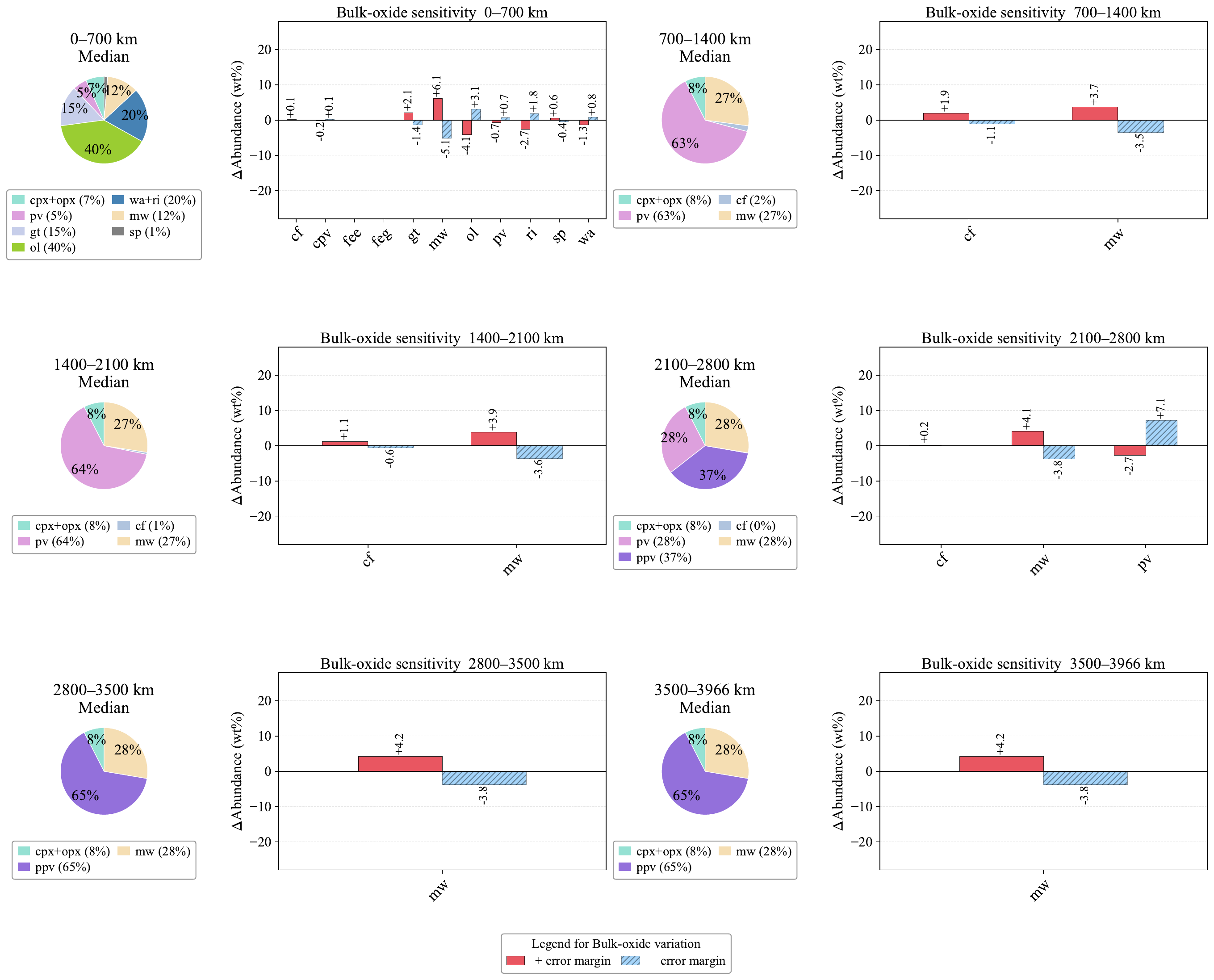}
    \caption{Sensitivity of mantle mineralogy to variations in composition corresponding to the upper and lower uncertainty bounds derived from the MvNA model. The data are presented in Table \ref{tab:bulk_composition}. The overall mantle mineralogy remains largely unchanged. Most variations are primarily confined to the lower mantle and transition zone, particularly in magnesiow\"ustite, bridgmanite (perovskite), and post-perovskite, and are smaller than those observed for the MvNP sensitivity.}
    \label{fig:MvNA_composition_sensitivity}
\end{figure*}

For the MvNA model, the overall sensitivity is smaller than that observed for the MvNP model.
As shown in the sensitivity plot in Figure \ref{fig:err_MvNA_phase}, the largest compositional variations for the MvNA model also occur in Mg, Fe, and Si.

Sections \ref{subsec:stellar_abundance_models}, \ref{subsec:inferring_bulk_composition}, and \ref{subsec:bulk_mantle} describe that the MvNA model is conditioned on elemental abundances inferred from stellar spectra, reflecting the limited availability of precise measurements for all elements.
Since the mean oxide abundances in the MvNA case differ from those in the MvNP case, the abundances of associated Fe--Mg silicate minerals vary for the uncertainty-bounded compositions, although their variations are less pronounced than in the corresponding MvNP results.

For this model, all variations remain within 8\% in both the upper and lower mantle.
In the upper mantle, the largest variations are observed for olivine, magnesiow\"ustite, garnet, ringwoodite and wadsleyite.
For the lower mantle, magnesiow\"ustite, bridgmanite (perovskite), and calcium-ferrite show significant variations in the modal abundance (see Figure \ref{fig:err_MvNP_phase}).

Overall, compositional uncertainties derived from the stellar abundance models produce variations that differ slightly in both magnitude and depth distribution between the two models.
The MvNA sensitivity is consistently smaller than that of the MvNP model, reflecting the difference in how each model is conditioned on stellar data. 
In neither case do the compositional perturbations alter the first-order mineralogical structure of the mantle predicted for GJ 486b by our ESH framework.

\section{Discussion} \label{sec:discussion}

The compositional characteristics and predicted mineralogy derived from our analysis of GJ 486b provide new insights into the diversity of interior structures among super-Earths and how they may differ from Earth's mantle regime.
Below, we present and discuss the possible mantle mineralogy of GJ 486b, its implications, and the limitations of our study.

\subsection{Mineralogical Regimes for GJ 486b} \label{subsec:mineralogy_GJ}

The mineralogical structure of GJ 486b reflects the combined effects of its mass, density, radius, internal thermal gradient, and Fe-rich mantle bulk composition. 
Although the overall sequence of silicate phase transitions broadly resembles that of Earth, the depth distribution, phase transition zones, and modal abundances of major mantle minerals differ systematically.


These differences arise primarily because pressure increases more rapidly with depth in GJ 486b as a consequence of its higher mass, larger radius, and stronger gravity. 
This leads to the stabilization of high-pressure phases at shallower depths and compresses mineral stability regimes relative to Earth \citep{sahu2025unveiling}.

Before discussing these regimes in detail, it is useful to briefly contextualize the results using the Earth validation presented in Section \ref{subsec:validation_HeFESTo}.
The comparison between the \texttt{HeFESTo}-only and ESH frameworks demonstrates that changes in bulk composition introduced by devolatilization lead to modest but systematic shifts in mineral stability regimes, while preserving the first-order mantle phase sequence \citep{Stixrude-II, wang2019volatility}. 
Both frameworks employ the same thermodynamic formulation to generate the phase diagram, so these differences primarily reflect variations in bulk composition rather than methodological effects. As shown in Table \ref{tab:bulk_composition}, the devolatilized Earth (BSE) composition shows relative enrichment in SiO$_2$, Al$_2$O$_3$, MgO, CaO, and Cr$_2$O$_3$  accompanied by slight depletion in FeO compared to the pre-devolatilization \texttt{HeFESTo}-Earth composition. 
This pattern reflects the preferential retention of refractory components and the relative depletion of moderately volatile and siderophile elements during planetary devolatilization.
Such oxide redistribution is consistent with the expected compositional effects of volatile loss \citep{McDonough1995}.
Also, Na$_2$O shows an apparent enrichment in devolatilized composition that is due to the fact that the \texttt{ExoInt} model reconstructs Earth's primitive mantle composition, predicting 0.42 wt\% Na$_2$O. This is in agreement with standard BSE empirical estimates (0.36 wt\%; \citep{McDonough1995}).
In contrast, for \texttt{HeFESTo} the mantle composition follows the depleted MORB mantle (DMM) estimate of \citet{workman2005major}, in which incompatible elements such as Na have been extracted into the continental crust over geologic time.
For the devolatilized Earth composition, we observe a slight narrowing of the olivine and bridgmanite stability zones, and slight broadening of the magnesiow\"ustite field.
These variations are consistent with experimental constraints on the sensitivity of mantle mineralogy to Fe/Mg ratios \citep{fei2004experimentally, murakami2004post, hirose2006postperovskite}.

The mineralogical differences between Earth before and after devolatilization are minor, whereas the differences between Earth and GJ 486b are significantly larger.
This contrast indicates that the mineralogy of GJ 486b is primarily governed by intrinsic planetary properties, including its bulk composition, $P$--$T$ profile and larger size and mass, rather than by modeling artifacts \citep{unterborn2019pressure}.

Figures \ref{fig:Earth_HeFESTo_devol} and \ref{fig:mvna_mvnp_final_gj} show that the upper mantle of GJ 486b is dominated by olivine, with subordinate pyroxenes, garnet, spinel, and magnesiow\"ustite.
While this assemblage is broadly similar to that of Earth, the relative stability regimes show modest differences.
Clinopyroxene occupy smaller stability range, and orthopyroxene became unstable, whereas the spinel regime is modestly expanded. 
Pyroxenes are particularly sensitive to both pressure amplification and Fe enrichment, further promoting their destabilization relative to olivine- and garnet-bearing assemblages.

The transition zone of GJ 486b preserves the typical olivine polymorphic sequence, with olivine transforming to wadsleyite and subsequently to ringwoodite \citep{kerschhofer1998polymorphic, wu2012experimental}.
Within this transition zone, the stability range and modal abundances of garnet and ringwoodite are reduced.
Although wadsleyite spans a depth range comparable to that on Earth, its modal abundance is lower, indicating that it is rapidly consumed by reactions leading to lower-mantle assemblages.
These trends indicate a preference for assemblages stable at higher pressures, leading to a reduced presence of intermediate-pressure phases \citep{umemoto2017phase}.
The stability of calcium-ferrite structured phases within the transition zone of GJ 486b, especially in the MvNP, further supports this interpretation \citep{Iskrina2022PostSpinel, Shim2000CaSiO3Pv}.
For ilmenite, both the MvNA and MvNP models show lower abundances relative to Earth's mantle; although the variation is noticeable in the MvNP model, it is negligible in the MvNA model. 
This variation reflects differences in Fe content, as the inferred mantle FeO content in GJ 486b is higher than Earth, and also the amount in the MvNP model is almost double that in the MvNA model.

Below the transition zone, the mantle of GJ 486b is dominated by bridgmanite, magnesiow\"ustite, calcium perovskite, and calcium-ferrite, as expected for massive rocky planets \citep{umemoto2017phase}. 
Although the phase identities resemble those of Earth's lower mantle, both the modal proportions and the relative abundances of cationic endmembers within these phases differ substantially, reflecting the combined effects of extreme pressure and variable Fe/Al content on phase relations \citep{katsura2025phase}.

In particular, the stability domain of bridgmanite is reduced under strong compression, while magnesiow\"ustite occupies a larger fraction of the lower mantle, consistent with theoretical models of super-Earth mineralogy \citep{umemoto2017phase}.
This redistribution of modal abundances arises from both Fe enrichment and elevated internal pressures and densities, which favor dense oxide and silicate structures.
 The stability of calcium-ferrite further indicates that aluminous and ferric components are increasingly accommodated within dense high-pressure crystal structures predicted at terapascal conditions \citep{HUANG2026100001}. 
Calcium perovskite remains stable over a depth range comparable to that on Earth but shows an increased modal abundance, consistent with enhanced Ca partitioning at high pressure.

At greater depths, bridgmanite transforms into post-perovskite, producing a lowermost mantle assemblage dominated by post-perovskite, magnesiow\"ustite, and calcium perovskite, as expected for super-Earth mantles \citep{umemoto2017phase}.
Although this transition also occurs in Earth's mantle, its expression in GJ 486b differs in both depth and extent. It occurs at an Earth-like depth in MvNA model but a slightly shallower depth in MvNP model. 
This shows that GJ 486b's transition zone depends on assumptions about its interior composition in the models. 
In addition, even though post-perovskite makes up a smaller fraction of the mantle at any given depth in GJ 486b than in Earth, due to the silica-limited bridgmanite field and the corresponding expansion of the magnesiow\"ustite field, its much thicker mantle means post-perovskite still occupies a larger portion of the lower mantle overall.

\subsection{Implications of Mineralogy on Mantle Physics} \label{subsec:Implications}

The mineralogical structure of GJ 486b has important implications for mantle physics and rheology.
The increased abundance of ferropericlase-rich assemblages suggests a mechanically weaker lower mantle compared to Earth, as ferropericlase is less viscous than bridgmanite-dominated assemblages \citep{yamazaki2001some}.
However, this weakening may be offset by the planet's high size and mass and strong gravitational gradient \citep{sahu2025unveiling}, which promotes gravitational stratification and may suppress large-scale convection.
At greater depths, the presence of bridgmanite--post-perovskite transitions implies strong depth-dependent variations in rheology and thermal conductivity, as post-perovskite is more anisotropic and thermally conductive than bridgmanite \citep{tsuchiya2004phase}.
These contrasts are likely to favor layered or multi-regime convection, with deep mantle regions partially decoupled from upper mantle circulation. 
The sharp density contrast at the core--mantle boundary (see Figure 4 in \citet{sahu2025unveiling}), together with widespread post-perovskite and Fe-rich silicates, suggests that heat transfer across the boundary may differ substantially from that of Earth.
While the high thermal conductivity of post-perovskite may enhance basal mantle heating, it may also stabilize long-lived basal layers and reduce core heat loss \citep{murakami2004post}.




\subsection{Limitations of the Study and Future Prospects} \label{subsec:Limitations}

The mineralogical structure of GJ 486b presented in this study is derived using a combination of modeling approaches, each of which involves inherent assumptions and limitations.
Accordingly, the predicted interior structure and mineralogy should be interpreted within the context of these methodological constraints.

The primary limitation arises from the limited availability and precision of elemental abundance measurements for M-dwarf host stars.
The cool photospheres of M-dwarfs produce complex spectra dominated by dense molecular bands (e.g., TiO, H$_2$O), which merge atomic lines together.
Obtaining high signal-to-noise ratio, high-resolution near-infrared (NIR) spectra from large telescopes is observationally challenging \citep{olander2025abundance, wang2024towards}.
As a result, direct and complete high-precision stellar abundance measurements are unavailable for many M-dwarf systems, including GJ 486.
To address this limitation, we adopt a population-based approach that infers chemically consistent stellar abundances using statistically representative compositions rather than system-specific measurements.
While this method captures the broad chemical diversity expected for M-dwarf hosts by correlating abundances with stellar parameters, it cannot fully account for star-to-star variations or unique chemical signatures \citep{Ting2012, Ness2015}.
Therefore, the inferred planetary bulk compositions should be interpreted as plausible estimates rather than definitive determinations of the true composition of GJ 486b.

The devolatilization model used in \texttt{ExoInt} to convert stellar abundances into planetary bulk compositions \citep{wang2019volatility, wang2019enhanced, wang2022detailed} introduces additional uncertainties.
This approach assumes that volatile loss during planet formation follows trends calibrated primarily on Solar System bodies.
It further assumes that the volatility-dependent elemental depletion observed for Earth represents a general pathway for rocky planet formation, despite known compositional variability among terrestrial planets within the Solar System.
The model does not explicitly simulate disk evolution, condensation-evaporation processes, planet migration, or accretionary impacts, and therefore captures broad compositional trends rather than system-specific formation histories.
Its predictions are sensitive to uncertainties in stellar abundance measurements and are limited to silicate-dominated compositions, making it less suitable for highly irradiated planets or systems with non-terrestrial chemistries \citep{wang2019volatility}.

The interior structure and mineralogical phase equilibria in this study also have inherent limitations.
These are computed using \texttt{SERPINT} \citep{sahu2025unveiling} to derive $P$--$T$ profiles and \texttt{HeFESTo} \citep{Stixrude-II} to calculate equilibrium mineral assemblages.
While these models provide a physically consistent framework, they rely on simplified representations of planetary interiors.
As discussed by \citet{Stixrude-II}, the thermodynamic formulations implemented in \texttt{HeFESTo} assume equilibrium phase relations and depend on parameterized equations of state calibrated primarily against experimental and theoretical constraints for Earth's mantle conditions.
At the extreme pressures and temperatures relevant to super-Earth interiors, direct experimental constraints on mineral stability, elasticity, and thermodynamic properties remain sparse, requiring extrapolation beyond well-constrained regimes \citep{Stixrude-II, murakami2004post}.
In addition, the pressure--temperature profile generated by \texttt{SERPINT} is based on one-dimensional interior structure and evolution models that assume laterally homogeneous material properties and parameterized convection \citep{sahu2025unveiling}.
As a result, uncertainties in mantle temperature, density, and composition propagate into the phase equilibria calculations performed by \texttt{HeFESTo}, influencing the predicted depths of phase transitions and the stability domains of minerals.

Our framework also does not capture three-dimensional mantle heterogeneity, dynamic phase disequilibrium, or tectonic regimes, all of which may influence mineralogical distributions in planetary interiors.

Despite these limitations, our approach provides a practical framework for exploring the mantle mineralogy of rocky exoplanets using host-star abundances together with fundamental planetary parameters and their associated sensitivities. Therefore, the results presented here should be interpreted as physically motivated first-order predictions of the interior mineralogy of GJ 486b. This framework can be extended to the broader sample of rocky exoplanets orbiting M-dwarfs as part of the ongoing JWST Rocky Worlds Director's Discretionary Time program, enabling the exploration of potential links between interior mineralogy and observed atmospheric properties or the presence of bare rocky worlds \citep{xue2025jwst}.

\section{Conclusion} \label{sec:conclusion}

This work presents a physically consistent framework that connects stellar abundances to the interior structure and mineralogy of rocky exoplanets. The framework integrates four key components: stellar abundance inference, devolatilization modeling to estimate planetary bulk and mantle compositions, interior structure calculations to derive pressure--temperature profiles, and thermodynamic equilibrium modeling to predict mantle mineralogy. The framework has been benchmarked against Earth and applied to the super-Earth GJ 486b, a target of the JWST Rocky Worlds Director's Discretionary Time program. The main findings of this study are summarized as follows:

\begin{enumerate}


     \item We demonstrate that stellar chemical abundances, when combined with physically motivated devolatilization models, provide robust constraints on planetary interior composition and mineralogy beyond those obtainable from mass and radius measurements alone.

    \item To enable this, we construct a stellar abundance model which was trained on the stellar data available on \textit{Gaia} DR3 and Hypatia catalogs. We use a multivariate Gaussian approach to predict the incomplete elemental abundances for M-dwarfs. We obtain a physically plausible bulk composition for GJ 486b, which is validated against that of Earth.

    \item By applying the devolatilization framework to solar abundances, we successfully reproduce the first-order mineralogical structure of Earth's mantle, including the major phase sequences and transition depths. This benchmark demonstrates the internal consistency of our coupled framework and supports its application to the exoplanet targets modeled in this study.

    \item Applying the framework to GJ~486b, we find that its mantle preserves the same sequence of major mineral phases as Earth's mantle but shows significant differences in modal phase abundances and transition depths, driven jointly by the planet's higher mass, steeper pressure gradient, and distinct bulk \text{Fe/Mg/Si} stoichiometry. As a consequence of the higher interior compression and distinct bulk composition, high-pressure silicate phases, such as bridgmanite, ferropericlase, and post-perovskite, stabilise at shallower physical depths, with phase transitions occurring over narrower depth intervals and post-perovskite persisting across a substantially larger fraction of the lower-mantle volume than on Earth. 



    
    \item Overall, the differences in mantle mineralogy between Earth and GJ~486b demonstrate that a planet's mineralogical structure depends strongly on its interior $P$--$T$ conditions and host-star-derived mantle composition.
\end{enumerate}

As observational constraints on exoplanet masses, radii, and host-star abundances continue to improve, the framework presented here provides a foundation for exploring the diversity of rocky planet interiors. It also enables investigations of potential links between interior mineralogy and atmospheric properties, particularly in the era of JWST observations of rocky exoplanets, including secondary-eclipse measurements of dayside thermal emission.

\section*{Acknowledgments}

L.M. acknowledges funding support from the DAE through the NISER project RNI 4011. L.M. also gratefully acknowledges support from Breakthrough Listen at the University of Oxford through a sub-award at NISER under Agreement R82799/CN013, provided as part of a global collaboration under the Breakthrough Listen project funded by the Breakthrough Prize Foundation. L.M. also extends thanks to Prof. Sumit Chakraborty from the Institut f\"ur Geologie, Mineralogie und Geophysik for the insightful discussions related to the interior chemistry of the Earth. We sincerely thank Dr. Claire Marie Guimond for the exceptionally thoughtful, constructive, and insightful review of this manuscript.
 



%

\setcounter{section}{0} 
\renewcommand{\thesection}{\Alph{section}} \refstepcounter{section}

\bibliographystyle{aasjournal}
\bibliography{references}

\end{document}